\documentclass[12pt]{article}

\usepackage[left=2.8cm,right=2.8cm,top=2.8cm,bottom=2.8cm]{geometry}

\usepackage[colorlinks=true,linkcolor=black,citecolor=black,urlcolor=blue,filecolor=black]{hyperref}

\let\OLDthebibliography\thebibliography
\renewcommand\thebibliography[1]{
  \OLDthebibliography{#1}
  \setlength{\parskip}{0pt}
  \setlength{\itemsep}{5.8pt plus 0.3ex}
}

\usepackage{amsmath}
\usepackage{amsfonts}
\usepackage{slashed}
\usepackage[nosort]{cite}
\usepackage[normalem]{ulem}

 \csname
@addtoreset\endcsname{equation}{section}

\def\a{\alpha}
\def\b{\beta}
\def\g{\gamma}
\def\G{\Gamma}
\def\d{\delta}
\def\D{\Delta}
\def\e{\epsilon}
\def\ve{\varepsilon}
\def\z{\zeta}
\def\h{\eta}
\def\th{\theta}
\def\Th{\Theta}

\def\l{\lambda}

\def\m{\mu}
\def\n{\nu}

\def\X{\Xi}

\def\r{\rho}

\def\o{\omega}
\def\O{\Omega}

\def\cF{{\cal F}}
\def\cG{{\cal G}}
\def\cH{{\cal H}}

\def\cJ{{\cal J}}

\def\cN{{\cal N}}
\def\cO{{\cal O}}

\def\cQ{{\cal Q}}

\def\be{\begin{equation}}
\def\ee{\end{equation}}
\def\bea{\begin{eqnarray}}
\def\eea{\end{eqnarray}}
\def\ba{\begin{array}}
\def\ea{\end{array}}

\def\nn{\nonumber}

\def\nd{\nabla\!\cdot}

\def\pe{\prime}
\def\12{\frac{1}{2}}
\def\pr{\partial}
\def\prd{\partial \cdot}

\newcommand{\bin}[2]{{#1 \choose #2}}

\declareslashed{}{\not}{-.55}{0}{\psi} 

\begin{document}


\vspace{30pt}

\begin{center}


{\Large\sc Higher-spin charges in Hamiltonian form \\[10pt]
II. Fermi fields} 


\vspace{25pt}
{\sc A.~Campoleoni${}^{\, a}$, M.~Henneaux${}^{\, a}$, S.~H\"ortner${}^{\, b}$ and A.~Leonard${}^{\, a,}$\footnote{Research Fellow of the Fund for Scientific Research-FNRS Belgium.}}

\vspace{10pt}
{${}^a$\sl\small
Universit{\'e} Libre de Bruxelles\\
and International Solvay Institutes\\
ULB-Campus Plaine CP231\\
1050 Brussels,\ Belgium
\vspace{10pt}

${}^b$\sl \small Centro de Estudios Cient{\'\i}ficos (CECs)\\ Casilla 1469, Valdivia, Chile\\
\vspace{10pt}

{\it andrea.campoleoni@ulb.ac.be, henneaux@ulb.ac.be,\\ hortner@cecs.cl, amaury.leonard@ulb.ac.be}

}

\vspace{40pt} {\sc\large Abstract} \end{center}

\noindent We build the asymptotic higher-spin charges associated with ``improper" gauge transformations for fermionic higher-spin gauge fields on Anti de Sitter backgrounds of arbitrary dimension.  This is achieved within the canonical formalism. We consider massless fields of spin $s+1/2$, described by a symmetric spinor-tensor of rank $s$ in the Fang-Fronsdal approach. We begin from a detailed analysis of the spin $5/2$ example, for which we cast the Fang-Fronsdal action in Hamiltonian form, we derive the charges and we propose boundary conditions on the canonical variables that secure their finiteness. We then extend the computation of charges and the characterisation of boundary conditions to arbitrary half-integer spin.  Our construction generalises to higher-spin fermionic gauge fields the known Hamiltonian derivation of supercharges in AdS supergravity.


\newpage


\tableofcontents


\section{Introduction}\label{sec:intro}

This paper is devoted to the construction of the conserved charges that are associated to the gauge symmetry entailed by massless fermionic higher-spin fields and that generalise the supercharge surface integrals of supergravity. We focus on Anti de Sitter (AdS) backgrounds and we employ the canonical approach to conserved charges developed for general relativity \cite{Regge} and already extended to spin $3/2$ gauge fields in the context of AdS (super)gravity \cite{AD1,HT}. The results reported in the following thus complement our companion study of fully-symmetric bosonic fields \cite{charges-bose}. The framework and the motivations for the study of surface charges in higher-spin theories were already illustrated in \cite{charges-bose}, so that here we confine ourselves to recall the key points of our strategy and to stress the new features introduced by Fermi fields. 

We compute charges starting from the Fang-Fronsdal action \cite{fang-fronsdal} on AdS backgrounds \cite{fronsdal-AdS}. This is the action describing the \emph{free} dynamics of massless fermions of spin $s+1/2$; nevertheless we expect that our final expression for the charges will continue to apply in the non-linear theory, at least in some regimes and for a relevant class of solutions. As discussed at length in \cite{charges-bose}, this expectation relies on the idea that asymptotically the fields become weak and the linearised theory applies. It is also supported by numerous examples of charges linear in the fields in gravitational theories (see e.g.~\cite{covariant-charges1}), although one should keep in mind that counterexamples exist as well. These typically involve scalar fields coupled to gravity for which the back reaction of the scalar field on the metric cannot be neglected even asymptotically \cite{scalar1,scalar2,scalar3,scalar4}. One may therefore foresee similar effects also in higher-spin theories, since the known non-linear models usually require matter fields in the spectrum for consistency (see e.g.\ the reviews~\cite{review-Vasiliev_1,review-Vasiliev_2}). Knowledge of the linearised charges is anyway a key input to tackle also the regimes where non-linear corrections become relevant and we defer to future studies the analysis of the possible impact of matter couplings. 

A similar perspective has been employed in a study \cite{ABJ-triality} of the asymptotic symmetries of supersymmetric Vasiliev's theories in four space-time dimensions \cite{vasiliev-susy_1,vasiliev-susy_2,vasiliev-susy_3,vasiliev-susy_4,vasiliev-susy_5,vasiliev-susy_6}, which have been inferred from the effect of vacuum-preserving gauge transformations on linearised solutions of the equations of motion. Our work complements these results --~obtained without any reference to an action principle~-- by providing a solid framework for the analysis of conserved charges and their associated asymptotic symmetries. One virtue of the Hamiltonian derivation of surface charges is indeed that these  generate asymptotic symmetries via the Dirac bracket. In the context of fermionic higher-spin theories, however, fully fledged investigations of this interplay between charges and asymptotic symmetries have been limited to three-dimensional examples \cite{susy_AdS-CFT_1,3d-fermi-asymptotics_1,3d-fermi-asymptotics_2,finite-superalgebras_1,finite-superalgebras_2,3d-fermi-asymptotics_3}. Aside from providing further tools for the study of asymptotic symmetries in supersymmetric higher-spin theories, our results can also accompany ``unfolding'' techniques \cite{unfolding-offshell_1,unfolded-charges1,unfolding-offshell_2,unfolded-charges2} in determining the fermionic charges of the proposed black hole solutions of Vasiliev's equations \cite{4D-BH1,4D-BH2,4D-BH3} or generalisations thereof.

Since we build charges within the canonical formalism, we first need to cast the Fang-Fronsdal action in Hamiltonian form. Given that the action is of first order in the derivatives, this step essentially requires to distinguish the dynamical variables from the Lagrange multipliers enforcing the first-class constraints that generate the gauge symmetry. In flat space this analysis has been performed in \cite{Hfermi1} for fields of arbitrary half-integer spins and revisited in the spin-$5/2$ case in \cite{Hfermi2} and more recently in \cite{hypergravity}. Hamiltonian actions involving a different field content --~inspired by the frame formulation of general relativity~-- have also been considered for higher-spin fermions in both flat \cite{Hfermi-frame_1} and (A)dS backgrounds \cite{Hfermi-frame_2}. Here we extend the presentation of the spin-$5/2$ action in \cite{Hfermi1,hypergravity} to AdS and (partly) to higher spins.

Once the constraints have been determined, we build the canonical generators of gauge transformations, which contain boundary terms. These boundary terms are non-vanishing in the case of ``improper gauge transformations''  \cite{Benguria:1976in} -- i.e., transformations that take the form of gauge transformations but produce a non-trivial effect on the physical system because they do not go to zero fast enough at infinity -- and are identified with the higher-spin surface charges. Improper gauge transformations are determined by boundary conformal Killing spinor-tensors (precisely defined in the text by definite equations), up to proper gauge transformations. Hence, to each conformal Killing spinor-tensor of the boundary is associated a well-defined higher-spin surface charge.

To illustrate the logic of the procedure, in sect.~\ref{sec:example} we first detail the rewriting in canonical form of the AdS Fang-Fronsdal action for a spin-5/2 Dirac field. We then provide boundary conditions on the dynamical variables and on the parameters of the gauge transformations preserving them. We finally use this information to evaluate the charges at spatial infinity and we conclude by discussing the peculiarities of the three-dimensional case. In sect.~\ref{sec:arbitrary} we move to arbitrary half-integer spin: first of all we observe that, for the sake of computing charges, it is sufficient to know the form of the constraints in flat space. We therefore bypass a detailed Hamiltonian analysis of the AdS theory and we build surface charges from flat-space constraints. We then present boundary conditions on the dynamical variables inspired by the behaviour at spatial infinity of the solutions of the free equations of motion (recalled in Appendix~\ref{app:boundary}) and we verify that they guarantee finiteness of the charges. We conclude with a summary of our results and with a number of appendices. Appendices~\ref{app:conventions} and \ref{app:first_order} recall our conventions and some useful facts about first-order actions, while Appendix~\ref{app:conformal-killing} discusses the conformal Killing spinor-tensors which play a crucial role in the study of charges and asymptotic symmetries.

\section{Spin-5/2 example}\label{sec:example}

To apply the techniques developed within the canonical formalism to compute surface charges, we begin by rewriting the AdS Fang-Fronsdal action in a manifestly Hamiltonian form. The charges are then identified with the boundary terms that enter the canonical generator of improper gauge transformations. Finally, we propose boundary conditions on the dynamical variables and on the deformation parameters that give finite asymptotic charges.

\subsection{Hamiltonian and constraints}\label{sec:example-H}

Our starting point is the Fang-Fronsdal action\footnote{Actions for spin-$5/2$ gauge fields in four dimensional Minkowski space have also been presented independently in \cite{schwinger,BvHvNdW}.} for a massless spin-5/2 Dirac field on  AdS$_d$ \cite{fronsdal-AdS}, described by a complex-valued spinor-tensor $\psi_{\m\n}^\a$ which is symmetric in its base-manifold indices $\m, \n$ (see also e.g.\ the review \cite{review-free}):\footnote{The sign in front of the mass-like term is conventional. One can change it provided one also changes the sign of the $L^{-1}$ terms in the gauge transformation \eqref{cov-var}, consistently with the option to send \mbox{$\g^\m \to - \g^\m$}. See sect.~\ref{sec:3d} for a discussion of the effect of this transformation (when $d=3$).}
\begin{align}
S & = -
i \!\int\! d^d x \sqrt{-\bar{g}} \left\{ \bar{\psi}^{\mu\nu} \slashed{D} \psi_{\mu\nu}
+ 2\, \bar{\psi}_{\m\n} \gamma^\n \gamma^\l \gamma^\r D_\l \psi_{\r}{}^\m - \frac{1}{2} \, \bar{\psi}\slashed{D}\psi 
+ \bar{\psi}_{\m\n} \g^\m D^\n \psi + \bar{\psi} D\!\cdot \slashed{\psi} \right. \nn \\
& - 2 \left( \bar{\psi}^{\m\n} D_{\m} \slashed{\psi}{}_\n + \bar{\psi}_{\m\n} \g^\m D\!\cdot \psi^\n \right) + \left. \frac{d}{2L} \left( \bar{\psi}^{\m\n} \psi_{\m\n} - 2\, \bar{\psi}^{\m}{}_{\n} \g^\n \g^\r \psi_{\m\r} - \frac{1}{2}\, \bar{\psi} \psi \right) \right\} . \label{action_lag_5_2}
\end{align}
Spinor indices will always be omitted, while $D$ stands for the AdS covariant derivative \eqref{cov-fermi}, $L$ is the AdS radius,\footnote{All results of this subsection apply also to de Sitter provided that one maps $L \to iL$.} slashed symbols denote contractions with $\g^\m$ and omitted indices signal a trace, so that e.g.\ $\slashed{\psi}{}_\m = \g^\n \psi_{\m\n}$ and $\psi = g^{\m\n} \psi_{\m\n}$. In the previous formulae we employed ``curved'' $\g$ matrices, which are related to ``flat'' ones as $\g^\m = e^\m{}_{\!A} \hat{\g}^A$, where $e^\m{}_{\!A}$ is the inverse vielbein. Our conventions are also detailed in Appendix \ref{app:conventions}. The action \eqref{action_lag_5_2} is invariant under the gauge transformations
\be \label{cov-var}
\d \psi_{\m\n} = 2 \left( D_{(\m} \e_{\n)} + \frac{1}{2L}\, \g_{(\m} \e_{\n)} \right)
\ee
generated by a $\g$-traceless spinor-vector. In \eqref{cov-var} and in the following, parentheses denote a symmetrisation of the indices they enclose and dividing by the number of terms in the sum is understood.

Being of first order, the action \eqref{action_lag_5_2} is almost already in canonical form. However, one would like to distinguish the actual phase-space variables from the Lagrange multipliers that enforce the first-class constraints associated to the gauge symmetry \eqref{cov-var}. In flat space the rewriting in canonical form of the Fang-Fronsdal action for a spin-5/2 Majorana field in $d = 4$ has been presented in \cite{Hfermi1,Hfermi2,hypergravity}; to extend it to Dirac fields on Anti de Sitter backgrounds of generic dimension, we parameterise the AdS metric with the static coordinates
\be \label{AdS}
ds^2 = - f^2(x^k) dt^2 + g_{ij}(x^k) dx^i dx^j \, .
\ee 
We also choose the local frame such that the non-vanishing components of the vielbein and the spin connection are
\be \label{local-frame}
e^0 = f dt \, , \quad e^i = e_j{}^i dx^j \, , \quad \o^{0i} = e^{ij} \pr_j f dt \, .
\ee
To separate dynamical variables and Lagrange multipliers within the components of $\psi_{\m\n}$, we recall that time derivatives of the gauge parameter can only appear in the gauge variations of Lagrange multipliers (see e.g.\ \cite{Fradkin:1977hw,Teitelboim:1977fs} and \S~3.2.2 of \cite{Hbook}). This criterion allows one to identify the dynamical variables with the spatial components $\psi_{ij}$ of the covariant field and with the combination
\be \label{def-xi}
\Xi = f^{-2} \psi_{00} - 2\, \g^0 \g^i \psi_{0i} \, .
\ee
The remaining components of $\psi_{\m\n}$ play instead the role of Lagrange multipliers. The covariant gauge variation \eqref{cov-var} breaks indeed into
\begin{subequations} \label{var-can}
\begin{align}
\d \psi_{ij} & = 2 \left( \nabla_{\!(i} \e_{j)} + \frac{1}{2L}\, \g_{(i} \e_{j)} \right) , \label{var-psi} \\
\d \Xi & = -\, 2\, \slashed{\nabla} \slashed{\e} + \frac{d+1}{L}\, \slashed{\e} \, , \label{var-xi} \\
\d \psi_{0i} & = \dot{\e}_i + f^2 \g^0 \left( \nabla_{\!i} \slashed{\e} - \frac{1}{2}\, \slashed{\G} \e_i - \G_i\, \slashed{\e} - \frac{1}{2L} \left( \e_i + \g_i \slashed{\e} \right) \right) ,
\end{align}
\end{subequations}
where we remark that Latin indices are restricted to spatial directions, while $\nabla$ is the covariant derivative for the spatial metric $g_{ij}$ and $\G_i$ denotes the Christoffel symbol $\G^0_{0i}$. The latter depends on $g_{00}$ as
\be
\G_i = f^{-1} \pr_i f \, .
\ee 
Moreover, from now on slashed symbols denote contractions involving only spatial indices, e.g.\ $\slashed{\e} = \g^i \e_i$.
The cancellation of time derivatives in \eqref{var-xi} follows from Fronsdal's constraint on the gauge parameter:\footnote{The rewriting of the gauge variations in \eqref{var-can} also relies on the identities $\o_t{}^{0k} \hat{\g}_k = f\, \G_k \g^k$ and \mbox{$\G^k_{00} = f^2 g^{kl} \G_l$}, that hold thanks to \eqref{AdS} and \eqref{local-frame}, while $\G^0_{00} = \G^0_{kl} = \G^k_{0l} = 0$.}
\be \label{gamma-constr}
\g^\m \e_\m = 0 \quad \Rightarrow \quad \e_0 = f^2 \g^0 \g^i \e_i \, .
\ee

The previous splitting of the fields is confirmed by the option to cast the action \eqref{action_lag_5_2} in the following canonical form:  
\be \label{can-action_5-2}
S = \!\int\! d^dx \left\{ \frac{1}{2}\! \left( \Psi^\dagger{}_{\!\!\!A}\, \o^{AB} \dot{\Psi}_B - \dot{\Psi}^\dagger{}_{\!\!\!A}\, \o^{AB} \Psi_B \right) - \psi^\dagger_{0k}\, \cF^k[\Psi] - \cF^\dagger_k[\Psi^\dagger]\, \psi_{0}{}^k - \cH[\Psi,\Psi^\dagger] \right\} ,
\ee
where we collected the phase-space variables by defining
\be \label{def_symplectic}
\Psi_A = 
\begin{pmatrix}
\psi_{kl} \\
\Xi
\end{pmatrix} , 
\qquad
\o^{AB} = 
\begin{pmatrix}
\o^{kl|mn} & \o^{kl|\bullet} \\
\o^{\bullet|mn} & \o^{\bullet|\bullet}
\end{pmatrix} .
\ee
The kinetic term is specified by the symplectic 2-form\footnote{With respect to Appendix~\ref{app:first_order} -- which recalls some general facts about first-order Grassmanian actions like \eqref{can-action_5-2} -- the symplectic 2-form has here implicit spinor indices and should incorporate a spatial delta function: $\O^{AB}\left(\vec{x},\vec{x}'\right) = \o^{AB} \d \left( \vec{x} - \vec{x}'\right)$.} $\o^{AB}$ with components 
\begin{subequations} \label{2_form_5_2}
\begin{align}
\o^{kl|mn} & = i\, \sqrt{g} \left( g^{k(m}g^{n)l} - 2\, \g^{(k} g^{l)(m} \g^{n)} - \frac{1}{2}\, g^{kl} g^{mn} \right) ,  \\
\o^{kl|\bullet} & = \o^{\bullet|kl} = -\, \frac{i}{2}\, \sqrt{g}\, g^{kl} \, , \\
\o^{\bullet|\bullet} & = \frac{i}{2}\,\sqrt{g} \, , 
\end{align}
\end{subequations}
where $\sqrt{g}$ involves only the determinant of the spatial metric. The symplectic 2-form satisfies $\o^{AB} = - \left( \o^{BA} \right)^{\dagger}$ and its inverse $\o_{AB}$ --~which enters the definition of the Dirac brackets given below~-- reads 
\begin{subequations} \label{inverse-omega}
\begin{align}
\o_{kl|mn} & = \frac{i}{\sqrt{g}} \left( -\, g_{k(m}g_{n)l} + \frac{2}{d}\, \g_{(k} g_{l)(m} \g_{n)} + \frac{1}{d}\, g_{kl} g_{mn} \right) , \\
\o_{kl|\bullet} & = \o_{\bullet|kl} = \frac{i}{d\sqrt{g}}\, g_{kl} \, , \\
\o_{\bullet|\bullet} & = -\, \frac{i}{\sqrt{g}}\, \frac{d+1}{d} \, .
\end{align}
\end{subequations}
The constraints enforced by the Lagrange multipliers $\psi_{0k}$ are instead
\be \label{constr}
\begin{split}
\cF_k = i\,\sqrt{g}\, \Big\{& 2 \left( \nd \psi_k - \g_k \nd \slashed{\psi} - \slashed{\nabla} \slashed{\psi}{}_k \right) - \nabla_{\!k} \psi + \g_k \slashed{\nabla} \psi - \nabla_{\!k} \X - \g_k \slashed{\nabla} \X \\
& + \frac{d}{2L} \left( 2\, \slashed{\psi}{}_k + \g_k \psi - \g_k \X \right) \Big\} \, ,
\end{split}
\ee
while the Hamiltonian reads 
\be \label{H}
\begin{split}
\cH & = i\,f\sqrt{g} \left\{ \left( \bar{\psi}_{kl} \slashed{\nabla} \psi^{kl} + \frac{1}{2}\, \G_m \bar{\psi}_{kl} \g^m \psi^{kl} \right) - \frac{3}{2} \left( \bar{\X} \slashed{\nabla} \X + \frac{1}{2}\, \G_k  \bar{\X} \g^k \X \right) \right. \\
& + 2 \left( \bar{\psi}_{kl} \g^k \g^m \g^n \nabla_{\!m} \psi_{n}{}^l + \frac{1}{2}\, \G_m \bar{\psi}_{kl} \g^k \g^m \g^n \psi_n{}^l \right) - \frac{1}{2} \left( \bar{\psi} \slashed{\nabla} \psi + \frac{1}{2}\, \G_k \bar{\psi} \g^k \psi \right) \\
& - 2 \left( \bar{\psi}^{kl} \nabla_{\!k} \slashed{\psi}{}_l + \bar{\psi}_{kl} \g^k \nd \psi^l + \G_m \bar{\psi}_{kl} \g^k \psi^{lm} \right) + \left( \bar{\psi}_{kl} \g^k \nabla^l \psi + \bar{\psi}\, \nd \slashed{\psi} + \G_k \bar{\psi} \slashed{\psi}{}^k \right) \\ 
& - \left( \bar{\X}\, \nd \slashed{\psi} + \bar{\psi}_{kl} \g^k \nabla^l \X + \G_k \bar{\X} \slashed{\psi}{}^k \right) + \frac{1}{2} \left( \bar{\X} \slashed{\nabla} \psi + \bar{\psi} \slashed{\nabla} \Xi + \G_k \bar{\Xi} \g^k \psi \right) \\ 
& \left. -\, \frac{3}{4}\, \G_k \left( \bar{\Xi} \g^k \psi - \bar{\psi} \g^k \X \right) + \frac{d}{4L} \left( 2\, \bar{\psi}_{kl} \psi^{kl} - 4\, \bar{\slashed{\psi}}{}_k \slashed{\psi}{}^k - \bar{\psi} \psi - 3\, \bar{\X} \X + \bar{\X} \psi + \bar{\psi} \X \right) \right\} .
\end{split}
\ee
Note that integrating by parts within $\cH$ generates contributions in $\G_k$ due to the overall dependence on $f(x^k)$. The terms collected within each couple of parentheses in \eqref{H} give an hermitian contribution to the action thanks to this mechanism. 

Following the steps outlined in Appendix~\ref{app:first_order} (to which we refer for more details), the knowledge of the symplectic 2-form allows to derive the Dirac brackets between fields:
\begin{subequations}\label{Dirac_5_2}
\begin{align}
\{ \psi^{\phantom{\dagger}}_{kl}(\vec{x}) , \psi^\dagger_{mn}(\vec{x}^{\,\pe}) \}_D & = \frac{i}{\sqrt{g}} \left( -\,g_{k(m} g_{n)l} + \frac{2}{d}\, \g_{(k} g_{l)(m} \g_{m)} + \frac{1}{d}\, g_{kl} g_{mn} \right) \d(\vec{x} - \vec{x}^{\,\pe}) \, , \\
\{ \psi^{\phantom{\dagger}}_{kl}(\vec{x}) , \Xi^\dagger(\vec{x}^{\,\pe}) \}_D & = \{ \Xi(\vec{x}) ,  \psi^\dagger_{kl}(\vec{x}^{\,\pe}) \}_D =    \frac{i}{d \sqrt{g}}\, g_{kl}\, \d(\vec{x} - \vec{x}^{\,\pe}) \, , \\
\{ \Xi(\vec{x}) , \Xi^\dagger(\vec{x}^{\,\pe})  \}_D & = -\,\frac{i}{\sqrt{g}}\, \frac{d+1}{d}\, \d(\vec{x} - \vec{x}^{\,\pe}) \, .
\end{align}
\end{subequations}
These are the same expressions as in flat space (compare e.g.\ with sect.~3 of \cite{hypergravity}). This was to be expected since the AdS action differs from the Minkowski one only through its mass term and its covariant derivatives (which are modified only by the addition of algebraic terms), neither of which modifies the kinetic term, containing one time derivative.

\subsection{Gauge transformations}\label{sec:example-gauge}

The action \eqref{action_lag_5_2} is invariant under \eqref{cov-var} for a $\g$-traceless $\epsilon_{\m}$ and this induces the variations \eqref{var-psi} and \eqref{var-xi} for the variables which are dynamically relevant. The constraints \eqref{constr} that we have just obtained are of first class on AdS (see \eqref{1st-class} below), and they generate these gauge transformations through their Dirac brackets with the fields.

The canonical generator of gauge transformations is indeed 
\begin{equation} \label{generator_5_2}
\mathcal{G} [ \lambda^k , \lambda^{\dagger\, l} ] = 
\int d^{d-1} x \left( \lambda^{\dagger\, k} \mathcal{F}_k + \mathcal{F}^{\dagger}{}_{\!k}\, \lambda^k \right) + Q[ \lambda^k , \lambda^{\dagger\, l} ] \, ,
\end{equation}
where $Q$ is the boundary term one has to add in order that $\mathcal{G}$ admit well defined functional derivatives, i.e.\ that its variation be again a bulk integral \cite{Benguria:1976in}:
\begin{equation} \label{deltaG}
\d \mathcal{G} = \int d^{d-1} x \left(\d \psi^{\dagger}{}_{\!\!kl}  A^{kl} + \d \Xi^{\dagger} B  + A^{\dagger kl} \d \psi_{kl} + B^{\dagger} \d \Xi \right) .
\end{equation}
The gauge variations of the dynamical variables are recovered from the Dirac brackets with the constraint, including its surface addition, as follows:
\begin{subequations} \label{var-bracket}
\begin{align}
\d \psi_{kl} &=
\left\lbrace \psi_{kl} , \mathcal{G} \right\rbrace_D = \omega_{kl\vert mn} A^{mn}  + \omega_{kl\vert \bullet} B \,, \\[5pt]
\d \Xi &= 
\left\lbrace \Xi , \mathcal{G} \right\rbrace_D = \omega_{\bullet\vert kl} A^{kl} + \omega_{\bullet\vert \bullet} B \,,
\end{align}
\end{subequations}
where the $\o_{AB}$ are the components of the inverse of the symplectic 2-form, given in \eqref{inverse-omega}.
Inserting into (\ref{generator_5_2}) the constraints (\ref{constr}), one obtains
\begin{align}
A^{kl} &= i  \sqrt{g} \left\lbrace
2 \left(
\nabla^{(k} \lambda^{l)} - \g^{(k} \nabla^{l)} \slashed{\lambda} - \g^{(k} \slashed{\nabla} \lambda^{l)}
\right) 
+ g^{kl}\! \left( \slashed{\nabla} \slashed{\lambda} - \nabla\!\cdot\!\lambda \right) - \frac{d}{2L}\! \left( 2\, \g^{(k} \lambda^{l)} + g^{kl} \slashed{\lambda} \right)
\right\rbrace , \nn \\
B &= i  \sqrt{g} \left\lbrace - \nabla \cdot \lambda - \slashed{\nabla} \slashed{\lambda} + \frac{d}{2L} \slashed{\lambda} \right\rbrace .
\end{align}
Substituting the values of $\o_{AB}$ from \eqref{inverse-omega}, one gets back the gauge transformations (\ref{var-can}) (for $\psi_{kl}$ and $\Xi$) with the identification $\lambda^k = \epsilon^k$. 

The variations \eqref{var-bracket} leave the constraints and the Hamiltonian invariant up to the constraints themselves:
\be \label{1st-class}
\d \cF_i = 0 \, , \qquad\qquad
\d \cH = - \left( \d \psi^\dagger_{0k} - \dot{\e}^\dagger_k\right) \cF^k - \cF^{\dagger\, k}_{\phantom{0k}} 
\left( \d \psi^{\phantom{\dagger}}_{0k} - \dot{\e}^{\phantom{\dagger}}_k \right) \, .
\ee
On the one hand, when combined with the variation of the kinetic term, these relations just reflect the gauge invariance of the Fang-Fronsdal action \eqref{action_lag_5_2} on AdS.\footnote{The variation of the constraints vanishes provided that the spatial metric be of constant curvature. To reproduce the variation of $\cH$ in \eqref{1st-class} one has instead to impose that the full space-time metric be of constant curvature.} On the other hand, given the link between gauge transformations and Dirac brackets recalled above, they also imply that both the constraints and the Hamiltonian are of first class and that there are no secondary constraints. This is confirmed by the associated counting of local degrees of freedom (see e.g.\ \S~1.4.2 of \cite{Hbook}):
\be
\#\, \textrm{d.o.f.} = 2^{\left[\frac{d}{2}\right]} \bigg(\! \underbrace{\frac{(d-1)d}{2} + 1}_{\textrm{dynamical variables}} -\ \,2 \!\!\!\!\!\!\underbrace{(d-1)\!\!\!\!\phantom{\frac{1}{2}}}_{\textrm{1st class constr.}} \!\!\!\!\bigg) = 2^{\left[\frac{d-2}{2}\right]} (d-3)(d-2) \, .
\ee
In $d=4$ the right-hand side is equal to four as expected, and in arbitrary $d$ it reproduces the number of degrees of freedom of a spin-5/2 Dirac fermion (compare e.g.\ with \cite{modave1}).

The boundary term $Q[\lambda^k , \lambda^{\dagger\, l}]$ will be of crucial importance in the following, since it gives the asymptotic charges. Its variation has to cancel the boundary terms generated by the integrations by parts putting the variation of $\cG[\lambda^k , \lambda^{\dagger\, l}]$ in the form \eqref{deltaG}. Being linear in the fields, these variations are integrable and yield:\footnote{Here $d^{d-2}S_k \equiv d^{d-2} x\, \hat{n}_k$, where $\hat{n}_k$ and $d^{d-2} x$ are respectively the normal and the product of differentials of the coordinates on the $d-2$ sphere at infinity (e.g.\ $d^{2} x = d\th d\phi$ for $d=4$).}
\be \label{charge_5-2}
\begin{split}
Q[\lambda^k , \lambda^{\dagger\, l}] = -
 i \int d^{d-2} S_k \sqrt{g}\,  \Big\lbrace
& 2\,  \lambda^{\dagger j} \psi_j{}^k - 2
\, \lambda^{\dagger}{}_{\!\!j} \g^j \g^l 
 \psi_l{}^k - 2\, \lambda^{\dagger j} \g^k \g^l \psi_{jl} 
- \lambda^{\dagger k} \psi \\  
& + \lambda^{\dagger}{}_{\!\!j} \g^j \g^k \psi
- \lambda^{\dagger k}\, \X - \lambda^{\dagger}{}_{\!\!j} \g^j \g^k\, \X  
\,\Big\rbrace + \textrm{h.c.}
\end{split}
\ee
In the definition of $Q$, we also adjusted the integration constant so that the charge vanishes for the zero solution. Note that, since the constraint \eqref{constr} contains a single derivative, the expression above for the boundary term on AdS is the same as that one we would have obtained in Minkowski. For clarity, in this example we displayed the complete Hamiltonian form of the AdS Fang-Fronsdal action; still knowledge of the constraints in flat space suffices to compute charges. 
In sect.~\ref{sec:arbitrary} we shall follow this shortcut when dealing with arbitrary half-integer spins.

\subsection{Boundary conditions}\label{sec:example-bnd}

The previous considerations remain a bit formal in the sense that the surface integrals (\ref{charge_5-2}) might diverge.  This is where boundary conditions become relevant.  In fact, for generic theories, the problem of cancelling the unwanted surface terms that appear in the variation of the Hamiltonian and the problem of defining boundary conditions are entangled and must be considered simultaneously, because it is only for some appropriate boundary conditions that the requested charges are integrable and that one can perform the cancellation. The reason why we got above (formal) integrability of the charges without having to discuss boundary conditions is that the constraints are linear.  One can then construct formal expressions for the charges first since integrability is automatic.

To go beyond this somewhat formal level and  to evaluate the asymptotic charges, however, we have to set boundary conditions on the dynamical variables. In analogy with the strategy we employed for Bose fields \cite{charges-bose}, we propose to use as boundary conditions the falloffs at spatial infinity of the solutions of the linearised field equations in a convenient gauge. We check in sect.~\ref{sec:example-charges} that these conditions make the charges finite.

In Appendix~\ref{app:boundary} we recall the behaviour at the boundary of AdS of the solutions of the Fang-Fronsdal equations of motion; in spite of being of first order, these equations admit two branches of solutions, related to different projections that one can impose asymptotically on the fields.\footnote{The existence of two branches of solutions associated to different projections on the boundary values of the fields is not a peculiarity of higher spins. For spin-$3/2$ fields on AdS it has been noticed already in \cite{HT}, while for spin-$1/2$ fields it has been discussed e.g.\ in \cite{AdS/CFT-spinors,boundary-dirac}.} In a coordinate system in which the AdS metric reads
\be \label{poincare}
ds^2 = \frac{dr^2}{r^2} + r^2\, \h_{IJ} dx^I dx^J \, ,
\ee
the solutions in the \emph{subleading branch} behave at spatial infinity ($r \to \infty$) as
\begin{subequations} \label{fall-off_5-2}
\begin{align}
\psi_{IJ} & = r^{\frac{5}{2}-d}\, \cQ_{IJ}(x^K) + \cO(r^{\frac{3}{2}-d}) \, , \\[5pt]
\psi_{rI} & = \cO(r^{-d-\frac{1}{2}}) \, , \\[5pt]
\psi_{rr} & = \cO(r^{-d-\frac{7}{2}}) \, .
\end{align}
\end{subequations}
We remark that capital Latin indices denote all directions which are transverse to the radial one (including time) and that here and in the following we set the AdS radius to $L = 1$. The field equations further impose that $\cQ_{IJ}$ satisfies the following conditions:
\begin{subequations} \label{bnd-constr_5-2}
\begin{align}
\pr^J \cQ_{IJ} = \hat{\g}^J \cQ_{IJ} = 0 \, , \label{div-gamma} \\[5pt]
\left( 1 + \hat{\g}^r \right) \cQ_{IJ} = 0 \, , \label{proj-Q}
\end{align}
\end{subequations}
where a hat indicates ``flat'' $\g$ matrices, that do not depend on the point where the expressions are evaluated. For instance, $\hat{\g}^r = \delta^r{}_{\!A} \hat{\g}^A$. Eqs.~\eqref{fall-off_5-2} and \eqref{bnd-constr_5-2} define our boundary conditions.

In the case of spin $3/2$ included in the discussion of sect.~\ref{sec:bnd}, the boundary conditions dictated by the subleading solution of the linearised e.o.m.\ agree with those considered for $\cN = 1$ AdS supergravity in four dimensions \cite{HT}.\footnote{This is actually true up to a partial gauge fixing allowing one to match the falloffs of the radial component. We refer to sect.~\ref{sec:bnd} for more details.} This theory is known in closed form, and finiteness of the charges and consistency have been completely checked. Moreover, the agreement in the spin-3/2 sector extends a similar matching between the subleading falloffs of linearised solutions and the boundary conditions generally considered in literature for gravity \cite{charges-bose}. These are our main motivations to adopt the boundary conditions defined by subleading linearised solutions for arbitrary values of the spin. See also sect.~\ref{sec:3d}, where we show how the conditions above allow one to match results previously obtained in the Chern-Simons formulation of three-dimensional higher-spin gauge theories \cite{susy_AdS-CFT_1,3d-fermi-asymptotics_1,3d-fermi-asymptotics_2,finite-superalgebras_1,finite-superalgebras_2,3d-fermi-asymptotics_3}.

Since the action is of first order, the constraints \eqref{constr} only depend on the dynamical variables $\psi_{ij}$ and $\Xi$, that somehow play both the role of coordinates and momenta. The boundary conditions \eqref{fall-off_5-2} and \eqref{bnd-constr_5-2}, obtained from the covariant field equations, can therefore be easily converted in boundary conditions on the canonical variables:
\begin{subequations} \label{Hboundary_5-2}
\begin{alignat}{5}
\psi_{\a\b} & = r^{\frac{5}{2}-d} \cQ_{\a\b} + \cO(r^{\frac{3}{2}-d})  \, , \qquad 
& \Xi & = -\, r^{\frac{1}{2}-d} \cQ_{00} + \cO(r^{-d-\frac{1}{2}}) \, , \\[5pt]
\psi_{r\a} & = \cO(r^{-d-\frac{1}{2}}) \, , \qquad 
& \psi_{rr} & = \cO(r^{-d-\frac{7}{2}}) \, .
\end{alignat}
\end{subequations}
In the formulae above we displayed explicitly only the terms which contribute to the charges (see sect.~\ref{sec:example-charges}) and we used Greek letters from the beginning of the alphabet to indicate the coordinates that parameterise the $d-2$ sphere at infinity. Furthermore, we used the $\g$-trace constraint \eqref{div-gamma} to fix the boundary value of $\Xi$.

\subsection{Asymptotic symmetries}\label{sec:example-symm}

In order to specify the deformation parameters that enter the charge \eqref{charge_5-2}, we now identify all gauge transformations preserving the boundary conditions of sect.~\ref{sec:example-bnd}. We begin by selecting covariant gauge transformations compatible with the fall-off conditions \eqref{fall-off_5-2}, and then we translate the result in the canonical language. 

Asymptotic symmetries contain at least gauge transformations leaving the vacuum solution $\psi_{\m\n} = 0$ invariant. These are generated by $\g$-traceless Killing spinor-tensors of the AdS background, which satisfy the conditions
\be \label{killing-ads}
D_{(\m} \e_{\n)} + \frac{1}{2}\, \g_{(\m} \e_{\n)} = 0 \, , \qquad\qquad
\g^\m \e_\m = 0 \, ,
\ee
and generalise the Killing spinors that are considered in supergravity theories (see \cite{AdS-killing-spinors} for a discussion of the Killing spinors of AdS$_d$ along the lines we shall follow for higher spins). We are not aware of any classification of $\g$-traceless Killing spinor-tensors of AdS spaces of arbitrary dimension, but they have been discussed for $d = 4$ \cite{ABJ-triality} and they are expected to be in one-to-one correspondence with the generators of the higher-spin superalgebras classified in \cite{HSsuperalgebras}. In sect.~\ref{app:conformal-killing} we shall also show that the number of independent solutions of \eqref{killing-ads} is the same as that of its flat limit, whose general solution is given by \eqref{sol-flat-killing}. These arguments indicate that --~as far as the free theory is concerned~-- non-trivial asymptotic symmetries exist in any space-time dimension and we are going to classify them from scratch in the current spin-$5/2$ example. Along the way we shall observe that, when $d > 3$, asymptotic and exact Killing spinor-tensors only differ in terms that do not contribute to surface charges.

To identify the gauge transformations that preserve the boundary conditions \eqref{fall-off_5-2}, one has to analyse separately the variations of components with different numbers of radial indices. In the coordinates \eqref{poincare}, if one fixes the local frame as
\be \label{localframe-text}
e_r{}^A = - \frac{1}{r}\, \d_r{}^A \, , \qquad 
e_I{}^J = \o_I{}^{rJ} = r\, \d_I{}^J \, , \qquad
\o_r^{\m\n} = \o_I{}^{JK} = 0 \, ,
\ee 
one obtains the conditions
\begin{align}
\d \psi_{IJ} & = 2 \left( \pr_{(I} \e_{J)} + \frac{r}{2}\, \hat{\g}_{(I|} \left( 1-\hat{\g}_r \right) \e_{|J)} \right) + 2\, r^3\, \h_{IJ} \e_r = \cO(r^{\frac{5}{2}-d}) \, , \label{dpsi-ij} \\[7pt]
\d \psi_{rI} & = \frac{1}{r} \left( r\pr_r - \frac{4+\hat{\g}_r}{2} \right) \e_I + \left( \pr_I + \frac{r}{2}\, \hat{\g}_I \left( 1 - \hat{\g}_r \right) \right) \e_r = \cO(r^{-d-\frac{1}{2}}) \, , \label{dpsi-ri} \\[5pt]
\d \psi_{rr} & = \frac{2}{r} \left( r\pr_r + \frac{2-\hat{\g}_r}{2} \right) \e_r = \cO(r^{-d-\frac{7}{2}}) \, . \label{dpsi-rr}
\end{align}
Fronsdal's $\g$-trace constraint $\slashed{\e} = 0$ implies instead
\be \label{gmu-emu}
\hat{\g}^r \e_r = r^{-2}\, \hat{\g}^I \e_I \, ,
\ee
thus showing that the radial component of the gauge parameter, $\e_r$, is not independent. It is anyway convenient to start analysing the conditions above from \eqref{dpsi-rr}, which is a homogeneous equation solved by
\be \label{e-r}
\e_r = r^{-\frac{1}{2}}\, \l^{+}(x^k) + r^{-\frac{3}{2}}\, \l^{-}(x^k) + \cO(r^{-d-\frac{5}{2}})\, , 
\quad \textrm{with} \quad \hat{\g}^{r} \l^\pm = \pm\, \l^\pm \, .
\ee
Substituting in \eqref{dpsi-ri} one obtains
\be \label{e-i}
\e_I = r^{\frac{5}{2}}\, \z^+_I + r^{\frac{3}{2}}\, \z^-_I + \frac{r^{\frac{1}{2}}}{2} \left( \pr_I \l^+ + \hat{\g}_{I} \l^- \right) + \frac{r^{-\frac{1}{2}}}{2}\, \pr_I \l^- + \cO(r^{\frac{1}{2}-d}) \, , 
\ee
where the new boundary spinor-vectors that specify the solution satisfy
\be \label{proj-z}
\hat{\g}^{r} \z_I^\pm = \pm\, \z_I^\pm \, .
\ee

The gauge parameter is further constrained by \eqref{dpsi-ij} and \eqref{gmu-emu}. The latter equation implies
\be \label{g-traces}
\slashed{\z}^+ \equiv \hat{\g}^I \z^+_I = 0 \, , \qquad\
\l^+ = \slashed{\z}^- \, ,
\ee
and the differential conditions
\be \label{diff-constr}
\slashed{\pr} \l^+ = - (d+1) \l^- \, , \qquad\
\slashed{\pr} \l^- = 0 \, . 
\ee
As shown in Appendix~\ref{app:independent}, the relations \eqref{diff-constr} are however not independent from the constraints imposed by \eqref{dpsi-ij}, so that we can ignore them for the time being. We stress that in the equations above and in the rest of this subsection, contractions and slashed symbols only involve sums over transverse indices and flat $\g$ matrices, so that \mbox{$\slashed{\pr} \l^\pm = \hat{\g}^I \pr_I \l^\pm$}.
Eq.~\eqref{dpsi-ij} implies instead
\be \label{dpsi-ij-exp}
\begin{split}
\d \psi_{IJ} & = 2\, r^{\frac{5}{2}} \left( \pr_{(I} \z_{J)}{}^{\!\!+} + \hat{\g}_{(I} \z_{J)}{}^{\!\!-} + \h_{IJ} \l^+ \right) + 2\, r^{\frac{3}{2}} \left( \pr_{(I} \z_{J)}{}^{\!\!-} + \h_{IJ} \l^- \right) \\[5pt]
& + r^{\frac{1}{2}} \left( \pr_I\pr_J \l^+ + 2\, \hat{\g}_{(I} \pr_{J)} \l^- \right) + r^{-\frac{1}{2}}\, \pr_I\pr_J \l^- = \cO(r^{\frac{5}{2}-d}) \, .
\end{split}
\ee
The cancellation of the two leading orders requires
\be
\l^+ = - \frac{1}{d-1} \left( \prd \z^+ + \slashed{\z}^- \right) =  - \frac{1}{d}\, \prd \z^+ \, , \qquad\
\l^- = - \frac{1}{d-1}\, \prd \z^- \, ,
\ee
plus the differential conditions presented below in \eqref{conformal-killing}. In Appendix~\ref{app:independent} we prove that these constraints also force the cancellation of the second line in \eqref{dpsi-ij-exp} when $d > 3$. At the end of this subsection we shall instead comment on how to interpret the additional terms that one encounters when $d=3$.

To summarise: parameterising the AdS$_d$ background as in \eqref{poincare} and fixing the local frame as in \eqref{localframe-text}, linearised covariant gauge transformations preserving the boundary conditions \eqref{fall-off_5-2} are generated by 
\begin{subequations} \label{final-gauge}
\begin{align}
\e^I & = r^{\frac{1}{2}}\, \z^{+I} + r^{-\frac{1}{2}}\, \z^{-I} - \frac{r^{-\frac{3}{2}}}{2d}\! \left( \pr^I \prd \z^+ + \frac{d}{d-1}\, \hat{\g}^{I} \prd \z^- \right) - \frac{r^{-\frac{5}{2}}}{2(d-1)}\, \pr^I \prd \z^- \quad \nn \\
& + \cO(r^{-\frac{3}{2}-d}) \, , \\[5pt]
\e^r & = - \frac{r^{\frac{3}{2}}}{d}\, \prd \z^+ - \frac{r^{\frac{1}{2}}}{d-1}\, \prd \z^- + \cO(r^{-d-\frac{1}{2}}) \, , \label{xir}
\end{align}
\end{subequations}
where the spinor-vectors $\z^\pm_I$ are subjected the chirality projections \eqref{proj-z} and satisfy\footnote{These conditions also allow \eqref{final-gauge} to satisfy the $\g$-trace constraint \eqref{gmu-emu}, which is not manifest in the parameterisation of the solution we have chosen.}
\begin{subequations} \label{conformal-killing}
\begin{align}
& \pr_{(I} \z_{J)}{}^{\!\!+} - \frac{1}{d-1}\,\h_{IJ}\, \prd \z^+ = -\, \hat{\g}_{(I} \z_{J)}{}^{\!\!-} + \frac{1}{d-1}\, \h_{IJ}\, \hat{\g}\cdot \z^- \, , \label{conf2+} \\
& \pr_{(I} \z_{J)}{}^{\!\!-} - \frac{1}{d-1}\,\h_{IJ}\, \prd \z^- = 0 \, , \label{conf2-}\\[5pt]
& \hat{\g}\cdot \z^+ = 0 \, , \label{constr2+} \\[4pt]
& \hat{\g}\cdot \z^- = - \frac{1}{d}\, \prd \z^+ \, . \label{constr2-}
\end{align}
\end{subequations}
The left-hand sides of \eqref{conf2+} and \eqref{conf2-} have the same structure as the bosonic conformal Killing-vector equation in $d-1$ space-time dimensions. For this reason we call here the solutions of \eqref{conformal-killing}  ``conformal Killing spinor-vectors''. When $d > 3$ there are \mbox{$2^{\left[\frac{d-1}{2}\right]}(d-2)(d+1)$} independent solutions that we display in Appendix~\ref{app:conformal-killing}, while when $d = 3$ the space of solutions actually becomes infinite dimensional (see sect.~\ref{sec:3d}). 

As discussed in sect.~\ref{sec:example-gauge}, Fang-Fronsdal's gauge parameters coincide with the deformation parameters entering the charge \eqref{charge_5-2}. Asymptotic symmetries are therefore generated by deformation parameters behaving as
\be \label{deform_5-2}
\l^\a = r^{\frac{1}{2}}\, \z^{+\a} + \cO(r^{-\frac{1}{2}}) \, , \qquad 
\l^r = \cO(r^{\frac{3}{2}}) \, .
\ee
As in \eqref{Hboundary_5-2}, Greek letters from the beginning of the alphabet denote coordinates on the $d - 2$ sphere at infinity and we specified only the terms that contribute to surface charges.

To conclude this subsection, we remark that an infinite number of solutions is not the unique peculiarity of the three-dimensional setup: in this case one indeed obtains
\be \label{3d-var}
\d \psi_{IJ} = - \frac{r^{-\frac{1}{2}}}{2}\, \pr_I \pr_J \prd \z^- + \cO(r^{-\frac{3}{2}}) 
\ee 
even considering gauge parameters that satisfy \eqref{final-gauge} and \eqref{conformal-killing}. One can deal with this variation in two ways: if one wants to solve the Killing equation \eqref{killing-ads}, one has to impose the cancellation of $\pr_I \pr_J \prd \z^-$ and the additional condition is satisfied only on a finite dimensional subspace of the solutions of \eqref{conformal-killing}. If one is instead interested only in preserving the boundary conditions \eqref{Hboundary_5-2}, which is the only option when the background is not exact AdS space, a shift of $\psi_{IJ}$ at $\cO(r^{-\frac{1}{2}})$ is allowed. In analogy with what happens for Bose fields \cite{metric3D,charges-bose}, the corresponding variation of the surface charges is at the origin of the central charge that appears in the algebra of asymptotic symmetries. In $d=3$ the spinor-vector $\z^{-I}$ entering the variation \eqref{3d-var} indeed depends on the spinor-vector $\z^{+I}$ entering the charges (see sect.~\ref{sec:3d}).

\subsection{Charges}\label{sec:example-charges}

Having proposed boundary conditions on both canonical variables (see \eqref{Hboundary_5-2}) and deformation parameters (see \eqref{deform_5-2}), we can finally obtain the asymptotic charges. In the coordinates \eqref{poincare}, the normal to the $d-2$ sphere at infinity is such that $\hat{n}_r = 1$ and $\hat{n}_\a = 0$. At the boundary the charge \eqref{charge_5-2} thus simplifies as
\be
\lim_{r\to\infty} Q[\lambda^k , \lambda^{\dagger\, l}]
 =  i \int d^{d-2} x \sqrt{g} \left\lbrace
2\, \lambda^{\dagger \alpha}\g^r \g^{\beta}\psi{}_{\alpha\beta}
- \lambda^{\dagger \alpha}\g_{\alpha} \g^r \left( \psi - \X  \right) \right\rbrace +\, \textrm{h.c.}
\ee
The terms which survive in the limit give a finite contribution to the charge; one can make this manifest by substituting their boundary values (where we drop the label $+$ on $\zeta^{+ I}$ to avoid confusion) so as to obtain 
\be \label{charge-cov}
Q = 2i \int d^{d-2} x  \left\lbrace
\bar{\zeta}^{I} \mathcal{Q}_{0I} + \bar{\mathcal{Q}}_{0I}\, \zeta^{I}
\right\rbrace .
\ee
This presentation of $Q$ relies on the $\g$-trace constraints on both $\mathcal{Q}_{IJ}$ and $\zeta^{I}$ and on the chirality conditions $(1+\hat{\g}^r) \cQ_{IJ} = 0$ and $(1-\hat{\g}^r) \z^I = 0$.
Remarkably, the result partly covariantises in the indices transverse to the radial direction.
The boundary charge thus obtained is manifestly conserved: it is the spatial integral of the time component of a conserved current since
\be
\cJ_I \equiv 2i\, \bar{\mathcal{Q}}_{IJ} \zeta^{ J} + \textrm{h.c.} \quad \Rightarrow \quad \prd \cJ = 2i \left( \pr^I
\bar{\cQ}_{IJ} \z^J + \bar{\cQ}_{IJ} \pr^{(I} \z^{J)} \right) + \textrm{h.c.} = 0 \, ,
\ee
where conservation holds thanks to \eqref{div-gamma} and \eqref{conf2+}. Eq.~\eqref{charge-cov} naturally extends the standard presentation of the bosonic global charges of the boundary theories entering the higher-spin realisations of the AdS/CFT correspondence (see e.g.\ sect.~2.5 of \cite{charges-bose}).

\subsection{Three space-time dimensions}\label{sec:3d}

We conclude our analysis of the spin-$5/2$ example by evaluating the charge \eqref{charge-cov} in three space-time dimensions, where several peculiarities emerge and we can compare our findings with the results obtained in the Chern-Simons formulation of supergravity \cite{CS-sugra1,CS-sugra2} and higher-spin theories \cite{susy_AdS-CFT_1,3d-fermi-asymptotics_1,3d-fermi-asymptotics_2,finite-superalgebras_1,finite-superalgebras_2,3d-fermi-asymptotics_3}. 

To proceed, it is convenient to introduce the light-cone coordinates $x^\pm = t \pm \phi$ on the boundary. Two inequivalent representations of the Clifford algebra, characterised by $\hat{\g}^0 \hat{\g}^1 = \pm \hat{\g}^2$, are available in $d=3$. Since we conventionally fixed e.g.\ the relative sign in the gauge variation \eqref{cov-var}, we shall analyse them separately by choosing
\be \label{clifford}
\hat{\g}^\pm = \begin{pmatrix}
0 & 2 \\
0 & 0 
\end{pmatrix} ,
\quad
\hat{\g}^\mp = \begin{pmatrix}
0 & 0 \\
-2 & 0 
\end{pmatrix} ,
\quad
\hat{\g}^r = \begin{pmatrix}
1 & 0 \\
0 & -1 
\end{pmatrix} .
\ee
Equivalently, one could fix the representation of the Clifford algebra once for all and analyse the effects of a simultaneous sign flip in the gauge variation \eqref{cov-var} and in the chirality projections \eqref{proj-Q} and \eqref{proj-z}.  

One can exhibit the peculiar form of surface charges in $d=3$ by studying the general solutions of the conditions \eqref{bnd-constr_5-2} and \eqref{conformal-killing} on $\cQ_{IJ}$ and $\z^I$. The constraints on $\cQ_{IJ}$ are solved by
\be
\cQ_{\mp\mp} = \begin{pmatrix}
0 \\
\cQ(x^\mp)
\end{pmatrix} ,
\qquad
\cQ_{+-} = \cQ_{\pm\pm} = \begin{pmatrix}
0 \\
0
\end{pmatrix} ,
\ee
where the signs are selected according to the conventions in \eqref{clifford}. Note that the divergence constraint is satisfied by suitable left or right-moving functions as for bosons, but the interplay between the $\g$-trace constraint \eqref{div-gamma} and the chirality projection \eqref{proj-Q} forces one of the two chiral functions to vanish. Similarly, the constraints on $\z^{\pm I}$ are solved by
\be
\z^{(+)\mp} = \begin{pmatrix}
\z(x^\mp) \\
0
\end{pmatrix} ,
\qquad
\z^{(-)\mp} = \frac{1}{3}\begin{pmatrix}
0 \\
\pr_\mp \z(x^\mp)
\end{pmatrix} ,
\qquad
\z^{(+)\pm} = \z^{(-)\pm} = \begin{pmatrix}
0 \\
0
\end{pmatrix} ,
\ee
where we enclosed between parentheses the label denoting the chirality, which appears in the covariant $\z^{\pm}_I$ of sect.~\ref{sec:example-symm}. 

When considering the representation of the Clifford algebra with $\hat{\g}^0 \hat{\g}^1 = \pm \hat{\g}^2$, the charge \eqref{charge-cov} therefore takes the form 
\be
Q_{d=3} = 2i \int d\phi\, \z^*(x^\mp) \cQ(x^\mp) + \textrm{c.c.} \, ,
\ee
so that a single Fang-Fronsdal field is associated to infinitely many asymptotic conserved charges, corresponding to the modes of an arbitrary function which is either left \emph{or} right-moving. On the contrary, a bosonic Fronsdal field is associated to both left \emph{and} right-moving charges \cite{metric3D,charges-bose}. From the Chern-Simons perspective, the counterpart of this observation is the option to define supergravity theories with different numbers of left and right supersymmetries \cite{extended-3d-sugra}. One can similarly define higher-spin gauge theories in $AdS_3$ by considering the difference of two Chern-Simons actions based on different supergroups (modulo some constraints on the bosonic subalgebra -- see e.g.\ sect.~2.1 of \cite{Wlambda}), while models with identical left and right sectors are associated to an even numbers of Fang-Fronsdal fields. 

\section{Arbitrary spin}\label{sec:arbitrary}

We are now going to generalise the results of the previous section to a spin $s + 1/2$ Dirac field, omitting the details that are not necessary to compute surface charges. As discussed at the end of sect.~\ref{sec:example-gauge}, these can be directly computed from the flat-space Hamiltonian constraints (which differ from the AdS$_d$ ones through algebraic terms) and the boundary conditions of fields and gauge parameters. We will focus on these two elements.

\subsection{Constraints and gauge transformations}\label{sec:gauge}

Our starting point is again the Fang-Fronsdal action for a massless spin $s + 1/2$ Dirac field on AdS$_d$ \cite{fronsdal-AdS}:
\be \begin{split} \label{fronsdal-action}
S = - i \int d^d x \sqrt{-\bar{g}}
& \left\{ \frac{1}{2}\,  \bar{\psi} \slashed{D} \psi
+ \frac{s}{2}\, \bar{\slashed{\psi}} \slashed{D} \slashed{\psi} - \frac{1}{4}\bin{s}{2} \bar{\psi}' \slashed{D} \psi'  
+ \bin{s}{2} \bar{\psi}' D\cdot\slashed{\psi} - s\, \bar{\psi} D \slashed{\psi} \right. \\
 & \left. +\, \frac{d+2(s-2)}{4L} \left( \bar{\psi} \psi - s\, \bar{\slashed{\psi}} \slashed{\psi} - \frac{1}{2}\bin{s}{2}\, \psi' \psi'
 \right) \right\} + \textrm{h.c.} 
\end{split}
\ee
The conventions are the same as in the spin-$5/2$ example (we will use the same static parametrisation of AdS$_d$, etc., from sect.~\ref{sec:example}), except that we will now leave all indices (tensorial and spinorial) implicit. The field is a complex-valued spinor-tensor \mbox{$\psi^{\a}_{\m_1 \cdots \m_s} = \psi^{\a}_{(\m_1 \cdots \m_s)}$}, which is fully symmetric in its $s$ base-manifold indices $\m_1 , \ldots , \m_s$. It will generally be denoted by $\psi$ and its successive traces will be indicated by primes or by an exponent in brackets: $\psi^{[k]}$ is the $k$th trace, while we often use $\psi'$ to denote a single trace.

The novelty of the general case is that there is an algebraic constraint on the field in addition to that on the gauge parameter: its triple $\g$-trace is required to vanish, $\slashed{\psi}{}^{\pe} = 0$. The only algebraically independent components of the field have therefore zero, one or two tensorial indices in the time direction: $\psi_{k_1 \cdots k_s}$, $\psi_{0 k_2 \cdots k_s}$ and $\psi_{00 k_3 \cdots k_s}$.

The action \eqref{fronsdal-action} is invariant under gauge transformations
\be \label{gauge-fermi}
\d \psi = s \left( D \e + \frac{1}{2L}\, \g\, \e \right)
\ee
generated by a $\g$-traceless spinor-tensor of rank $s-1$. In \eqref{gauge-fermi} and in the following, a symmetrisation of all free indices is implicit, and dividing by the number of terms in the sum is understood.

Similarly to the spin-$5/2$ analysis, we note that the action \eqref{fronsdal-action} is almost already in canonical form. To identify the Lagrange multipliers which enforce the first-class constraints associated to the gauge symmetry \eqref{gauge-fermi}, one can again select combinations whose gauge variation contains time derivatives of the gauge parameter. This leads to identify the dynamical variables with the spatial components $\psi_{k_1 \ldots k_s}$ of the covariant field and with the combination 
\be
\Xi_{k_1 \cdots k_{s-2}} = f^{-2} \psi_{00 k_1 \cdots k_{s-2}} - 2\, \g^0 \g^j \psi_{0 j k_1 \cdots k_{s-2}} \, .
\ee
The remaining independent components of the covariant fields, $\psi_{0 k_1 \cdots k_{s-1}} \equiv N_{k_1 \cdots k_{s-1}}$, play instead the role of Lagrange multipliers. The covariant gauge variation \eqref{gauge-fermi} breaks indeed into (with all contractions and omitted free indices being from now on purely spatial):
\begin{subequations}
\begin{align} 
\d \psi &=
s \left( \nabla \epsilon + \frac{1}{2L}\, \gamma \epsilon \right) , \label{spin_s_non_cov_gauge_transf}
\\
\d \Xi &= -\, 2 \slashed{\nabla} \slashed{\epsilon} - \left(s - 2\right) \nabla \epsilon' + \frac{1}{2L} \left[\, 2 \left(d + 1 + 2 \left(s - 2\right)\right) \slashed{\epsilon} - \left(s - 2\right) \g \e' \,\right] , \label{xi_non_cov_gauge_transf}
\\
\d N &=
\dot{\e} + f^2 \g^0 \left[ \left(s - 1\right) \nabla \slashed{\e} - \frac{1}{2}\, \slashed{\G} \e - \left(s - 1\right) \G \, \slashed{\e} - \frac{1}{2L} \left( \e + \left(s - 1\right) \g \slashed{\e} \right) \right] .
\end{align}
\end{subequations}
This choice of variables is further confirmed by injecting it back into the action \eqref{fronsdal-action}, and using the Fronsdal constraint that sets the triple $\g$-trace of the field to zero, which leads to the following identities 
\begin{subequations}
\begin{align}
\psi_{0 \cdots 0 k_{2n+1} \cdots k_s} &=
f^{2n} \left[ n \, \Xi^{[n-1]}_{k_{2n+1} \cdots k_s}
+ 2n \, \g^0 \slashed{N}^{[n-1]}_{k_{2n+1} \cdots k_s}
- \left(n-1\right) \psi^{[n]}_{k_{2n+1} \cdots k_s}\right] ,
\\
\psi_{0 \cdots 0 k_{2n+2} \cdots k_s} &=
f^{2n} \left[ n \, \g^0  \slashed{\Xi}^{[n-1]}_{k_{2n+2} \cdots k_s}
 +  \left(2n + 1\right)  N^{[n]}_{k_{2n+2} \cdots k_s}
 -  n \, \g^0  \slashed{\psi}^{[n]}_{k_{2n+2} \cdots k_s}\right] .
\end{align}
\end{subequations}
This brings the action into the canonical form 
\be \label{can_action_s}
S = \!\int\! d^dx \left\{ \frac{1}{2}\! \left( \Psi^\dagger{}_{\!\!\!A}\, \o^{AB} \dot{\Psi}_B - \dot{\Psi}^\dagger{}_{\!\!\!A}\, \o^{AB} \Psi_B \right) - N^\dagger \cF[\Psi] - \cF^\dagger[\Psi^\dagger]\, N - \cH[\Psi,\Psi^\dagger] \right\} ,
\ee
where we collected the phase-space variables by defining
\be \label{def_symplectic_s}
\Psi_A = 
\begin{pmatrix}
\psi_{k_1 \ldots k_s} \\
\Xi_{k_3 \ldots k_s}
\end{pmatrix} , 
\qquad
\o^{AB} = 
\begin{pmatrix}
\o^{k_1 \cdots k_s| l_1 \cdots l_s} & \o^{k_1 \cdots k_s |\bullet\, i_3 \cdots i_s} \\
\o^{\bullet j_3 \ldots j_s | l_1 \ldots l_s} & \o^{\bullet j_3 \ldots j_s|\bullet\, i_3 \ldots i_s}
\end{pmatrix} .
\ee
We will not exhibit all terms (symplectic 2-form, Hamiltonian, etc.) of this action, but only those which are necessary to compute surface charges, that is the constraints. These have $s-1$ implicit spatial indices symmetrised with weight one and read
\be \label{F_s}
\begin{split}
\mathcal{F} & =
-i\, \frac{\sqrt{g}}{2} \sum_{n = 0}^{[s/2]} \bin{s}{2n} \bigg\lbrace
2n\, \g \,  g^{n-1} \Big[\, 
\slashed{\nabla} \Xi^{[n-1]} 
+ \left(s - 2n\right) \nabla \slashed{\Xi}^{[n-1]}
+ 2\left(n - 1\right) \nabla\!\cdot \slashed{\Xi}^{[n-2]} \\ 
& 
- \slashed{\nabla} \psi^{[n]}
+ \left(s-2n\right) \nabla \slashed{\psi}^{[n]}
+ 2n\, \nabla \cdot \slashed{\psi}^{[n-1]}
\Big] +
\left(s-2n\right)  g^{n} \Big[\,  2n\, \nabla \!\cdot \Xi^{[n-1]} \\
& +\left(s - 2n - 1\right) \nabla \Xi^{[n]}
+ 2\left(n - 1\right) \nabla\!\cdot \psi^{[n]} + \left(s-2n - 1\right) \nabla \psi^{[n+1]}
+ 2\, \slashed{\nabla} \slashed{\psi}^{[n]}
\Big]
\bigg\rbrace + \cdots \!\!\!\!
\end{split}
\ee
The dots stand for algebraic contributions coming from the mass term in the action and possible contributions in $\G_i$, which do not contribute to the surface charge.

Through their Dirac brackets (built from the inverse of the symplectic 2-form), these constraints generate the gauge transformations \eqref{spin_s_non_cov_gauge_transf} and \eqref{xi_non_cov_gauge_transf}, under which the Hamiltonian and the constraints are invariant (which confirms that they are first class). The canonical generator of gauge transformations is again
\begin{equation} \label{generator_s}
\mathcal{G} [ \lambda , \lambda^{\dagger} ] = 
\int d^{d-1} x \left( \lambda^{\dagger} \mathcal{F} + \mathcal{F}^{\dagger} \lambda \right) + Q[ \lambda , \lambda^{\dagger} ] \, ,
\end{equation}
where $Q$ is the boundary term one has to add in order that $\mathcal{G}$ admits well defined functional derivatives, i.e.\ that its variation be again a bulk integral:
\begin{equation} \label{deltaG_s}
\d \mathcal{G} = \int d^{d-1} x \left(\d \psi^{\dagger}{}  A + \d \Xi^{\dagger} B  + A^{\dagger} \d \psi + B^{\dagger} \d \Xi \right) .
\end{equation}

To compute surface charges we are only interested in $Q$ (whose expression is independent of the terms we omitted in the constraint \eqref{F_s}). Its variation has to cancel the boundary terms generated by the integrations by parts putting the variation of $\cG$ in the form \eqref{deltaG_s}. Being linear in the fields, these variations are integrable and yield (we display explicitly the index $k$ contracted with $d^{d-2}S_k$.):
\be \label{charge_s}
\begin{split}
& Q[\lambda , \lambda^{\dagger}] = \frac{i}{2} \!\int\! d^{d-2} S_k \sqrt{g} \sum_{n = 0}^{[s/2]} \bin{s}{2n}
\bigg\lbrace 
2n  \Big[\, 
\slashed{\lambda}^{\dagger [n-1]}  \g^k \left( \Xi^{[n-1]} - \psi^{[n]} \right) \\ 
& + \left(s-2n\right) \slashed{\lambda}^{\dagger [n-1] k} \left( \slashed{\psi}^{[n]} + \slashed{\Xi}^{[n-1]} \right)
+ 2\left(n - 1\right) \slashed{\lambda}^{\dagger [n-1]} \slashed{\Xi}^{[n-2]  k}  
+ 2n\, \slashed{\lambda}^{\dagger [n-1]} \slashed{\psi}^{[n-1] k}
\Big] \\ 
& +
\left(s-2n\right) \Big[  \left(s - 2n - 1\right)  \lambda^{\dagger [n]k} \left( \Xi^{[n]} + \psi^{[n+1]} \right)
+ 2n\, \lambda^{\dagger [n]}\Xi^{[n-1]k}
\\
& + 2 \left(n - 1\right)\lambda^{\dagger [n]} \psi^{[n] k} 
+ 2\, \lambda^{\dagger [n]}\g^k \slashed{\psi}^{[n]}
\Big]
\bigg\rbrace + \textrm{h.c.}  \end{split}
\ee
In the definition of $Q$, we again adjusted the integration constant so that the charge vanishes for the zero solution.

\subsection{Boundary conditions and asymptotic symmetries}\label{sec:bnd}

As in the spin-$5/2$ example, we derive boundary conditions on the dynamical variables from the falloff at spatial infinity of the solutions of the Fang-Fronsdal equations in a convenient gauge, adopting the subleading branch. As shown in Appendix \ref{app:boundary}, with the parameterisation \eqref{local-frame} of the local frame the relevant solutions behave at spatial infinity ($r \to \infty$) as
\begin{subequations} \label{cond_field_s}
\begin{align}
\psi_{I_1 \cdots I_s} & = r^{\frac{5}{2} - d} \cQ_{I_1 \cdots I_s}(x^M) + \cO(r^{\frac{3}{2}-d}) \, , \\[5pt]
\psi_{r \cdots r I_1 \cdots I_{s-n}} & = \cO(r^{\frac{5}{2} - d - 3n}) \, , \label{radial-psi}
\end{align}
\end{subequations}
where capital Latin indices denote directions transverse to the radial one as in sect.~\ref{sec:example-bnd}. From now on we also set again $L = 1$. The boundary spinor-tensor $\cQ_{I_1 \cdots I_s}$ is fully symmetric as the Fang-Fronsdal field and satisfies
\begin{subequations} \label{bnd-constr}
\begin{align}
\prd \cQ = \slashed{\cQ} = 0 \, , \label{constr-s} \\[5pt]
\left( 1 + \hat{\g}^r \right) \cQ = 0 \, , \label{projQ-s}
\end{align}
\end{subequations}
where we omitted free transverse indices. Eqs.~\eqref{cond_field_s} and \eqref{bnd-constr} define our boundary conditions. For $s=1$ and $d=4$ the requirements on $\psi_I$ agree with those proposed for non-linear $\cN = 1$ supergravity in eq.~(V.1) of \cite{HT}. Our $\psi_r$ decays instead faster at infinity, but the leading term that we miss in \eqref{radial-psi} can be eliminated using the residual gauge freedom parameterised by the function $a(t,\th,\phi)$ in eq.~(V.5) of \cite{HT}. In conclusion, on a gravitino our boundary conditions agree with those considered in non-linear supergravity up to a partial gauge fixing that does not affect the charges.

The covariant boundary conditions \eqref{cond_field_s} fix the behaviour at spatial infinity of the dynamical variables as
\begin{subequations} \label{canonical-bnd}
\begin{align}
\psi_{\a_1 \cdots \a_s} & = r^{\frac{5}{2} - d} \cQ_{\a_1 \cdots \a_s} + \cO(r^{\frac{3}{2}-d}) \, , \\
\psi_{r \cdots r \a_1 \cdots \a_{s-n}} & = \cO(r^{\frac{5}{2} - d - 3n}) \, , \\[5pt]
\Xi_{\a_1 \cdots \a_{s-2}} & = -\, r^{\frac{1}{2}-d} \cQ_{00\a_1 \cdots \a_{s-2}} + \cO(r^{-d-\frac{1}{2}}) \, , \\
\Xi_{r \cdots r \a_1 \cdots \a_{s-n-2}} & = \cO(r^{\frac{1}{2} - d - 3n}) \, ,
\end{align}
\end{subequations}
where Greek indices from the beginning of the alphabet denote angular coordinates in the $d-2$ sphere at infinity. Moreover, we displayed only the dependence on the boundary values of the fields in the terms that actually contribute to surface charges.

The next step in the procedure we illustrated in sect.~\ref{sec:example} requires to identify all gauge transformations that do not spoil the boundary conditions \eqref{canonical-bnd}. We are now going to provide necessary conditions for the preservation of the asymptotic form of the fields, which generalise those given for $s=2$ in \eqref{final-gauge} and \eqref{conformal-killing}. A proof that they are also sufficient (along the lines of the proof presented for $s=2$ in Appendix~\ref{app:independent}) will be given elsewhere. We stress, however, that the rank-$s$ counterparts of \eqref{final-gauge} and \eqref{conformal-killing} also characterise the exact $\g$-traceless Killing spinor-tensors of AdS$_d$, which satisfy
\be \label{killing-ads-gen}
D \e + \frac{1}{2}\, \g\, \e = 0 \, , \qquad \slashed{\e} = 0 \, . 
\ee
The general solution of these equations is provided in Appendix~\ref{app:conformal-killing} for $s = 2$. It shows that the number of independent solutions is the same as in the flat-space limit, where Killing spinor-tensors are easily obtained (see \eqref{killing-flat}). In the following we assume that this concurrence holds for arbitrary values of the spin and, hence, that the conditions we are going to present admit as many independent solutions as integration constants in \eqref{killing-flat}.

To characterise the gauge parameters which generate asymptotic symmetries, one has to analyse separately the variations of components with different numbers of radial indices. We continue to omit transverse indices and we denote them as
\be
\psi_n \equiv \psi_{r \cdots r I_1 \cdots I_{s-n}} \, .
\ee
Similarly, we denote by $\e_n$ the component of the gauge parameter with $n$ radial indices. With the choice \eqref{local-frame} for the local frame, the variations of the field components must satisfy
\be \label{var-psi-s}
\begin{split}
\d \psi_n & = \frac{n}{r} \left( r\pr_r - \frac{2(2s-3n+1)+ \hat{\g}^r}{2} \right) \e_{n-1} + (s-n) \left( \pr + \frac{r}{2}\, \hat{\g}\, (1 - \hat{\g}^r ) \right) \e_n \\
& + 2 \bin{s-n}{2} \, r^3 \h\, \e_{n+1} = \cO(r^{\frac{5}{2} - d - 3n}) \, .
\end{split}
\ee
The constraint $\slashed{\psi}{}^{\pe} = 0$ actually implies that one can focus only on the variations of $\psi_0$, $\psi_1$ and $\psi_2$, since all other components are not independent. One also has to consider the constraint $\slashed{\e} = 0$, which implies
\be \label{gamma-trace-symm}
\hat{\g}^r \e_{n+1} = r^{-2} \slashed{\e}{}_{n}
\ee
and shows that the only independent component of the gauge parameter is the purely transverse one, that is $\e_0 \equiv \e_{I_1 \cdots I_{s-1}}$.

The equations \eqref{var-psi-s} require\footnote{This can be shown e.g.\ by considering the redundant variation of $\psi_s \equiv \psi_{r \cdots r}$, which gives a homogeneous equation for $\e_{s-1} \equiv \e_{r \cdots r}$ as in sect.~\ref{sec:example-symm}. One can then fix recursively the $r$-dependence of all other components of the gauge parameter.}
\be \label{gauge0-gen}
\e_0 = r^{2(s-1)} \left( r^{\frac{1}{2}} \z^+ + r^{-\frac{1}{2}} \z^- \right) + \sum_{k=1}^{s-1} r^{2(s-k-1)} \left(  r^{\frac{1}{2}} \a_k + r^{-\frac{1}{2}} \b_k \right) + \cO(r^{\frac{1}{2}-d})
\ee
together with the chirality projections
\be
(1\mp\hat{\g}^r) \z^\pm = 0
\ee
and similar restrictions on the subleading components: $(1-\hat{\g}^r)\a_k = (1+\hat{\g}^r)\b_k = 0$. The $\g$-trace constraint on the gauge parameter is then satisfied by
\begin{subequations}
\begin{align}
\e_{2n} = (-1)^n \sum_{k=n}^{s-n-1} r^{2(s-2n-k-1)} \left(  r^{\frac{1}{2}} \a_k^{[n]} + r^{-\frac{1}{2}} \b_k^{[n]} \right) + \cO(r^{\frac{1}{2}-d-6n}) \, , \\
\e_{2n+1} = (-1)^n \sum_{k=n}^{s-n-2} r^{2(s-2n-k)-5} \left(  r^{\frac{1}{2}} \slashed{\b}_k^{[n]} - r^{-\frac{1}{2}} \slashed{\a}_k^{[n]} \right) + \cO(r^{-d-\frac{5}{2}-6n}) \, .
\end{align}
\end{subequations}
Substituting these expressions in \eqref{var-psi-s} one obtains
\begin{subequations} \label{varpsi-gen}
\begin{align}
\d \psi_0 & = s\, r^{2(s-1)} \left\{ r^{\frac{1}{2}} \left[ \pr \z^+ + \g \z^- + (s-1)\, \h\, \slashed{\z}^{-} \right] + r^{-\frac{1}{2}} \left[ \pr \z^- - (s-1)\, \h\, \slashed{\a}{}_1 \right] \right\} \nn \\ 
& \!\!\!+ s \sum_{k=1}^{s-1} r^{2(s-k-1)}\! \left\{ r^{\frac{1}{2}}\! \left[ \pr \a_k + \g \b_k + (s-1)\, \h\, \slashed{\b}{}_k \right] + r^{-\frac{1}{2}}\! \left[ \pr \b_k - (s-1)\, \h\, \slashed{\a}_{k+1} \right] \right\} , \label{varpsi_0} \\[5pt]
\d \psi_1 & = \sum_{k=1}^{s-1} r^{2(s-k)-3}\! \left\{ r^{\frac{1}{2}}\! \left[ -2k\,\a_k + (s-1) \left( \pr \slashed{\b}{}_{k-1} - \g \slashed{\a}{}_k \right) - (s-1)(s-2)\, \h\, \a^{\pe}_k \right] \right. \nn \\
& \!\!\!\left. +\, r^{-\frac{1}{2}}\! \left[ -2k\, \b_k - (s-1) \left( \pr \slashed{\a}_k + (s-2) \h\, \b^{\,\pe}_k \right) \right] \right\} . \label{varpsi_1}
\end{align}
\end{subequations}
In complete analogy with the analysis of sect.~\ref{sec:example-symm}, the cancellation of the first line in \eqref{varpsi_0} requires
\begin{subequations} \label{generic-conf-kill}
\begin{align}
\pr \z^+ - \frac{s-1}{d+2s-4}\, \h\, \prd \z^+ + \hat{\g}\, \z^- & = 0 \, , \label{conf1} \\[5pt]
\pr \z^- - \frac{s-1}{d+2s-5}\, \h\, \prd \z^- & = 0 \, , \label{conf2} \\[5pt]
\hat{\g}\cdot \z^+ & = 0 \, , \label{conf3} \\[5pt]
\hat{\g}\cdot\z^- + \frac{1}{d+2s-4}\, \prd\z^+ & = 0 \, . \label{conf4}
\end{align}
\end{subequations}
These conditions generalise the conformal Killing equations \eqref{conformal-killing} to arbitrary values of the rank. Their general solution is given below in \eqref{sol3d-killing} when $d = 3$, while for $d > 3$ it will be given elsewhere. Still, as anticipated, the detailed analysis of the $s=2$ case presented in Appendix~\ref{app:conformal-killing} makes us confident that the equations \eqref{generic-conf-kill} admit a number of independent solutions equal to the number of integration constants in \eqref{killing-flat}.

The subleading orders in \eqref{varpsi-gen} allow instead to fix the subleading components of the gauge parameter in terms of $\z^\pm$. For instance, one can manipulate these expressions to obtain the recursion relations presented in Appendix~\ref{app:independent}.
Let us stress that, in analogy with what we observed for $s = 2$, \eqref{varpsi_0} and \eqref{varpsi_1} provide a set of equations that are compatible only if one takes into account the constraints \eqref{generic-conf-kill}. We do not have yet a proof of the latter statement, but the analysis given for $s=2$ in Appendix~\ref{app:conformal-killing} gives strong indications that this is a robust assumption.

The deformation parameters that generate gauge transformations preserving the boundary conditions can then be related to Fang-Fronsdal's gauge parameters by comparing the Lagrangian field equations with their rewriting in \eqref{eomH}. In particular, the Dirac brackets with the constraints can be inferred from the terms with Lagrange multipliers contained in the equations expressing the time derivatives of the dynamical variables in terms of the spatial derivatives of $\psi_{\m\n}$. This shows that the gauge parameter can be identified with the canonical deformation parameter also for arbitrary values of the spin. As a result, the latter behaves at spatial infinity as
\be \label{def-par}
\l^{\a_1 \cdots \a_{s-1}} = r^{\frac{1}{2}} \z^{+\a_1 \cdots \a_{s-1}} + \cO(r^{-\frac{1}{2}}) \, , \qquad
\l^{r \cdots r \a_1 \cdots \a_{s-n-1}} = \cO(r^{\frac{1}{2}+n}) \, ,
\ee
where $\z^+$ satisfies the conformal Killing equations \eqref{generic-conf-kill} and, as in \eqref{canonical-bnd}, Greek indices from the beginning of the alphabet denote angular coordinates in the $d-2$ sphere at infinity.

\subsection{Charges}\label{sec:charges}

Having proposed boundary conditions on both canonical variables (see \eqref{canonical-bnd}) and deformation parameters (see \eqref{def-par}), we can finally evaluate the asymptotic charges. The charge \eqref{charge_s} simplifies at the boundary as
\be \label{asymptotic-charge}
\begin{split}
\lim_{r\to\infty} Q[\lambda , \lambda^{\dagger}] =
i \!\int\! d^{d-2} x \sqrt{g}\, 
\sum_{n = 0}^{[s/2]} 
\bin{s}{2n} \Big\lbrace & n\, \slashed{\lambda}^{\dagger [n-1]} \g^r \left( \Xi^{[n-1]} - \psi^{[n]} \right) \\
& + \left(s-2n\right) \lambda^{\dagger [n]}\g^r \slashed{\psi}^{[n]} \Big\rbrace + \textrm{h.c.} \, ,
\end{split}
\ee
where all implicit indices are now purely spatial and transverse and we dropped the label $+$ on $\zeta^{+}$ to avoid confusion.

The terms which survive in the limit give a finite contribution to the charge; one can make this manifest by substituting their boundary values so as to obtain 
\be \label{charge-cov-gen}
Q = s\,i \int d^{d-2} x  \left\lbrace
\bar{\zeta}^{I_2 \ldots I_s} \mathcal{Q}_{0 I_2 \ldots I_s} + \bar{\mathcal{Q}}_{0 I_2 \ldots I_s}\, \zeta^{I_2 \ldots I_s}
\right\rbrace ,
\ee
where we used the $\g$-trace constraints on both $\mathcal{Q}$ and $\zeta$ and the chirality conditions $(1+\hat{\g}^r) \cQ = 0$ and $(1-\hat{\g}^r) \z = 0$.
The result partly covariantises in the indices transverse to the radial direction as in the spin-$5/2$ example, thus making the conservation of the charge manifest. It is indeed the spatial integral of the current \mbox{$\cJ_I = \bar{\zeta}^{K_2 \ldots K_{s}} \mathcal{Q}_{I K_2 \ldots K_s} + \textrm{h.c.}$}, which is conserved thanks to \eqref{constr-s} and \eqref{generic-conf-kill}.

In three space-time dimensions the charge \eqref{charge-cov-gen} is actually given by a left or right-moving function also for arbitrary half-integer values of the spin. As in sect.~\ref{sec:3d}, one can deal with the two inequivalent representations of the Clifford algebra by choosing the $\g$ matrices as in \eqref{clifford}. The constraints on $\cQ$ are then solved by
\be
\cQ_{\mp \cdots \mp} = \begin{pmatrix}
0 \\
\cQ(x^\mp)
\end{pmatrix} ,
\qquad 
\cQ_{\pm \cdots \pm} = \cQ_{+ \cdots + - \cdots -} = \begin{pmatrix}
0 \\
0
\end{pmatrix} ,
\ee
where different signs correspond to the choices $\g^0 \g^1 = \pm \g^2$. Similarly, the conformal Killing equations \eqref{generic-conf-kill} are solved by
\begin{subequations} \label{sol3d-killing}
\begin{align}
\z^{(+)\mp \cdots \mp} & = \begin{pmatrix}
\z(x^\mp) \\
0
\end{pmatrix} ,
\qquad
\z^{(-)\mp \cdots \mp} = \frac{1}{2s-1}\begin{pmatrix}
0 \\
\pr_\mp \z(x^\mp)
\end{pmatrix} , \\
\z^{(\pm)\pm \cdots \pm} & = \z^{(\pm) + \cdots + - \cdots -} = \begin{pmatrix}
0 \\
0
\end{pmatrix} ,
\end{align}
\end{subequations}
where we enclosed again between parentheses the label denoting the chirality of the spinor-tensors. The charge \eqref{charge-cov-gen} takes therefore the form
\be
Q_{d=3} = s\, i \int d\phi\, \z^*(x^\mp) \cQ(x^\mp) + \textrm{c.c.} \, ,
\ee
which generalises the result for the spin-$5/2$ case discussed in sect.~\ref{sec:3d}.

\section{Conclusions}\label{sec:conclusions}

We have explicitly constructed higher-spin charges for fermionic gauge fields of arbitrary spin on AdS backgrounds in any number of spacetime dimensions, extending our analogous work on bosonic gauge fields \cite{charges-bose}. We have followed Hamiltonian methods. The charges appear as the surface integrals that must be added to the terms proportional to the constraints in order to make the generators of gauge transformations well-defined as phase-space generators. These integrals are finite with the boundary conditions that we have given, which crucially involve chirality-type projections generalising those of \cite{HT}. Improper gauge transformations --~associated to non-vanishing surface integrals~-- are determined by conformal Killing spinor-tensors of the boundary, and the corresponding charges take a simple, boundary-covariant expression in terms of them -- even though the intermediate computations are sometimes rather involved. While bosonic higher-spin charges have been discussed also following other approaches \cite{HScharges,unfolded-charges2}, to our knowledge our treatment provides the first presentation of fermionic higher-spin charges that applies to any number of space-time dimensions.

We confined our analysis to the linearised theory, which suffices to derive the charges. In this context, however, the charges are abelian and their Dirac brackets vanish (modulo possible central extensions in $d=3$). To uncover a non abelian algebra, one must evaluate the brackets in the non-linear theory, since the bulk terms do play a role in that computation. A similar situation occurs for Yang-Mills gauge theories, where the surface terms giving the charges coincide with those of the abelian theory. The non-abelian structure appears when one computes the algebra of the charges \cite{Abbott:1982jh}, a step which involves the full theory (or, generically, at least the first non-linear corrections in a weak field expansion \cite{metric3D}). 

By working within the linearised theory, we have been able to associate conserved charges to any gauge field of given spin, although the spectra of interacting higher-spin theories are typically very constrained. We also worked with Dirac fields, that can be defined for any $d$, but the Majorana and/or Weyl projections that one may need to consistently switch on interactions can be easily implemented in our approach. In general, we have not found obstructions to define non-trivial higher-spin charges in any number of space-time dimensions and for arbitrary multiplicities of any value of the spin, consistently with the chance to define higher-spin algebras with fermionic generators in any dimension \cite{4dHSsuperalgebras,HSsuperalgebras}. Possible constraints could emerge from interactions, but  let us point out that once one starts considering half-integer higher spins, more exotic options than standard supersymmetry may become available. For instance, one can define higher-spin theories with increasing multiplicities for the fermionic fields without introducing any obvious pathology --~apart from difficulties in identifying a superconformal subalgebra within their algebra of asymptotic symmetries (see e.g.\ \cite{ABJ-triality,3d-fermi-asymptotics_3})~-- or try to define ``hypersymmetric'' theories (see e.g.\ \cite{hypergravity}) --~with fermionic gauge symmetries but without any gravitino at all. To analyse these phenomena it will be very interesting to combine our current results with those obtained for Bose fields \cite{charges-bose}, and to analyse the effect of interactions on the algebra of surface charges, e.g.\ introducing them perturbatively in a weak field expansion (see \cite{cubic-symmetries-1,cubic-symmetries-2,metric3D} for related work restricted to bosonic models).

\subsection*{Acknowledgements}

We are grateful to G.~Barnich, X.~Bekaert, N.~Boulanger, G.~Lucena G\'omez, P.~Somberg, P.~Sundell and M.~Taronna for useful discussions. A.C.\ acknowledge the Institute for Theoretical Physics of ETH Zurich for hospitality and support while the manuscript was in preparation. This work was partially supported by the ERC Advanced Grant  ``High-Spin-Grav'' and by FNRS-Belgium (convention FRFC PDR T.1025.14, convention IISN 4.4503.15 and the grant FC-98386 for A.C.). The work of S.H. has been supported by the Fondecyt grant N\textordmasculine \; 3160781. The Centro de Estudios Cient\'ificos (CECs) is funded by the Chilean Government through the Centers of Excellence Base Financing Program of Conicyt.


\begin{appendix}

\section{Notation and conventions}\label{app:conventions}

We adopt the mostly-plus signature for the space-time metric $g_{\m\n}$ and we often distinguish among time and spatial components by breaking space-time indices as $\m = (0,i)$. Tangent-space indices are collectively denoted by capital Latin letters, but when we separate time and spatial directions we use the same letters as for the indices on the base manifold, i.e.\ $A = (0,i)$. The $\g$ matrices then satisfy
\be
\{ \hat{\g}^A , \hat{\g}^B \} = 2\, \h^{AB} \, , \quad
(\hat{\g}^0)^\dagger = -\, \hat{\g}^0 \, , \quad
(\hat{\g}^i)^\dagger = \hat{\g}^i \, ,
\ee
where the hat differentiates them from their curved counterparts involving the inverse vielbein:
\be
\g^\m = e^\m{}_{\!A}\, \hat{\g}^A \, .
\ee
For instance, with the choice \eqref{local-frame} for the local frame one has $\hat{\g}^0 = f \g^0$. Similarly, the Dirac conjugate is defined as $\bar{\psi} = \psi^\dagger \hat{\g}^0$, while the $\g$ matrices displayed explicitly in the Fronsdal action (see e.g.\ \eqref{action_lag_5_2} or \eqref{fronsdal-action}) are curved ones. 

The space-time covariant derivative acts on a spin $s+1/2$ field as
\be \label{cov-fermi}
D_{\r} \psi_{\m_1 \cdots \m_s} = \pr_\r \psi_{\m_1 \cdots \m_s} + \frac{1}{8}\, \o_\r{}^{AB} [ \hat{\g}_A , \hat{\g}_B ] \psi_{\m_1 \cdots \m_s} - s\, \G^\l{}_{\r(\m_1} \psi_{\m_2 \cdots \m_s)\l} \, ,
\ee
and it satisfies $D_\m \g_\n = 0$. In the definition we omitted spinor indices as in the rest of the paper. Moreover, indices between parentheses are meant to be symmetrised with weight one, i.e.\ one divides the symmetrised expression by the number of terms that appears in it. The spatial covariant derivative $\nabla$ is defined exactly as in \eqref{cov-fermi}, but with indices constrained to take values only along spatial directions. It also satisfies $\nabla_{\!i} \g_j = 0$.

On an (A)dS background the commutator of two covariant derivatives reads
\be
[ D_\m , D_\n ] \psi_{\r_1 \cdots \r_s} = \frac{s}{L^2} \left( g_{\n(\r_1} \psi_{\r_2 \cdots \r_s)\m} - g_{\m(\r_1} \psi_{\r_2 \cdots \r_s)\n} \right) - \frac{1}{2L^2}\, \g_{\m\n} \psi_{\r_1 \cdots \r_s} \, ,
\ee
where $\g_{\m\n} = \frac{1}{2} [ \g_\m , \g_\n ]$. This relation defines the (A)dS radius $L$ and suffices to fix the mass term in the Fronsdal action \eqref{fronsdal-action}.

When expanding tensors in components, we actually distinguish among four types of indices, depending on whether the time and/or radial coordinates are included or not. Greek letters from the middle of the alphabet include all coordinates, small Latin letters include all coordinates except $t$, capital Latin letters include all coordinates except $r$, while Greek letters from the beginning of the alphabet denote the angular coordinates on the unit $d-2$ sphere. In summary:
\begin{alignat}{3}
\m,\n,\ldots & \in \{t,r,\phi^1,\ldots,\phi^{d-2}\} \, , \qquad
& i,j,\ldots & \in \{r,\phi^1,\ldots,\phi^{d-2}\} \, , \nn \\
I,J,\ldots & \in \{t,\phi^1,\ldots,\phi^{d-2}\} \, , \qquad
& \a,\b,\ldots & \in \{\phi^1,\ldots,\phi^{d-2}\} \, .
\end{alignat}

Slashed symbols always denote a contraction with a $\g$ matrix, whose precise meaning depends on the context: the contraction may be with the full $\g^\m$ or with its spatial counterpart $\g^i$. In sect.~\ref{sec:example} omitted indices denote a trace that, similarly, may result from a contraction with the full space-time metric $g_{\m\n}$ or with the spatial metric $g_{ij}$. In most of sect.~\ref{sec:arbitrary} we omit instead all indices, which are always assumed to be fully symmetrised according to the conventions given above. Traces are instead denoted by an exponent between square brackets, so that, for instance,
\be \label{notation}
\psi^{[n]} \equiv \psi_{\m_1 \cdots \m_{s-2n}\l_1 \cdots \l_n}{}^{\!\l_1 \cdots \l_n} \, , \qquad
D \psi \equiv D_{(\m_1} \psi_{\m_2 \cdots \m_{s+1})} \, , \qquad
\g\, \psi \equiv \g_{(\m_1} \psi_{\m_2 \cdots \m_{s+1})} \, . 
\ee
In Appendix~\ref{app:boundary} we reinstate indices with the following convention: repeated covariant or contravariant indices denote a symmetrisation, while a couple of identical covariant and contravariant indices denotes as usual a contraction. Moreover, the indices carried by a tensor are substituted by a single label with a subscript indicating their total number. For instance, the combinations in \eqref{notation} may also be presented as
\be \label{example-repeated}
\psi^{[n]} = \psi_{\m_{s-2n}} \, , \qquad
D \psi = D_{\m} \psi_{\m_s} \, , \qquad
\g\,\psi \equiv \g_\m \psi_{\m_{s}} \, . 
\ee

\section{First-order Grassmannian actions}\label{app:first_order}

Let us consider an action of the form
\begin{equation} \label{action_first_order}
S \left[\Psi , \bar{\Psi}\right] = \int dt \left\lbrace
\theta^A\!\left( \bar{\Psi} \right) \dot{\Psi}_A + \dot{\bar{\Psi}}_A\, \bar{\theta}^A\!\left(\Psi\right) - H\left(\Psi , \bar{\Psi}\right)
\right\rbrace ,
\end{equation}
\noindent where the $\Psi_A$ are complex Grassmann variables (of which all indices are incorporated into a capital Latin letter).

Its variation is given by
\begin{equation}
\delta S = \int dt \left\lbrace
\delta \bar{\Psi}_A \left[ \Omega^{AB} \dot{\Psi}_B \, - \, \frac{\partial^L H}{\partial \bar{\Psi}_A} \right]
\, + \, \left[ \dot{\bar{\Psi}}_A \bar{\Omega}^{AB} \, - \, \frac{\partial^R H}{\partial \Psi_B}\right] \delta \Psi_B 
\right\rbrace ,
\end{equation}
\noindent where
\be
\Omega^{AB} =
\frac{\partial^L \theta^B}{\partial \bar{\Psi}_A}
\, - \, \frac{\partial^R \bar{\theta}^A}{\partial \Psi_B} \, ,
\qquad
\bar{\Omega}^{AB} =
 \frac{\partial^R \bar{\theta}^A}{\partial \Psi_B}
\, - \, \frac{\partial^L \theta^B}{\partial \bar{\Psi}_A} \, .
\ee
Note that we have $\Omega^{AB} = \left(\bar{\Omega}^{BA}\right)^{\ast} = -\, \bar{\Omega}^{AB}$. We assume that $\Omega^{AB}$ is invertible.  It is then called the symplectic 2-form.

If we define the inverse of this 2-form as $\Omega^{AB} \Omega_{BC} = \delta^A_C$ (and a similar relation holds for its conjugate), the equations of motion become
\begin{eqnarray}
\dot{\Psi}_A &=&
\Omega_{AB}\, \frac{\partial^L H}{\partial \bar{\Psi}_B} \, ,
\\
\dot{\bar{\Psi}}_A &=&
\frac{\partial^R H}{\partial \Psi_B}\, \bar{\Omega}_{BA} \, .
\end{eqnarray}
This suggests to define a Dirac bracket such that $\dot{F} = \left\lbrace F , H \right\rbrace_D$. It is
\begin{equation}
\left\lbrace F , G \right\rbrace_D =
\frac{\partial^R F}{\partial \Psi_A} \,\Omega_{AB}\, \frac{\partial^L G}{\partial \bar{\Psi}_B}
\ + \ \frac{\partial^R G}{\partial \Psi_A} \,\bar{\Omega}_{AB}\, \frac{\partial^L F}{\partial \bar{\Psi}_B} \, .
\end{equation}
It is antisymmetric and satisfies $\left\lbrace \Psi_A , \bar{\Psi}_B \right\rbrace_D = \Omega_{AB}$, the other brackets vanishing identically. 
These brackets would in fact appear as Dirac brackets had one introduced conjugate momenta for the variables $\Psi_A$ and eliminated the corresponding second class constraints that express these momenta in terms of the $\Psi_A$ through the Dirac bracket procedure.  This short-cut of the orthodox Dirac method is well known and mentioned  e.g.\ in~\cite{Hbook}.

 A term of the form $- \int dt \left\lbrace \bar{u}_a\, f^a\!\left[ \Psi \right] + \bar{f}^a \!\left[ \bar{\Psi} \right] u_a \right\rbrace$ added to the action (\ref{action_first_order}), with Lagrange multipliers $u_a$ and first-class 
 $f^a$, 
 $\bar{f}^a$, 
 will generate gauge transformations according to 
\be
\d \Psi_A = \left\lbrace \Psi_A , \bar{f}^a\! \left[ \bar{\Psi} \right] u_a \right\rbrace_D
=  \O_{AB}\, \frac{\partial^L \bar{f}^a}{\partial \bar{\Psi}_B}\, u_a \, .
\ee
These gauge transformations can also be read directly from the components of the Lagrangian equations of motion containing time derivatives of the dynamical variables, through the identification
\be \label{eomH}
\dot{\Psi}_A = \{ \Psi_A , H \}_D + \{ \Psi_A , \bar{f}^a\!\left[\bar{\Psi}\right] \}\, u_a \, .
\ee

If the kinetic term of the action is quadratic as in \eqref{can-action_5-2}, the ``momenta" $\theta^A$ are linear in the positions $\Psi_A$ and we have:
\begin{eqnarray}
\theta_A &=&
\frac{1}{2} \ \bar{\Psi}_B \Omega^{BA} \, ,
\\
\bar{\theta}_A &=& \frac{1}{2} \ \bar{\Omega}^{AB} \Psi_B 
\ = \ - \ \frac{1}{2} \ \Omega^{AB} \Psi_B \, ,
\end{eqnarray}
showing that the $\Psi$'s and the $\bar{\Psi}$'s are conjugate.

\section{Covariant boundary conditions}\label{app:boundary}

In this appendix we recall the falloff at the boundary of $AdS_d$ of the solutions of the Fang-Fronsdal equations of motion (see also \cite{shadows-fermi}). To achieve this goal we partially fix the gauge freedom, and we also exhibit the falloffs of the parameters of the residual gauge symmetry (which include the $\g$-traceless Killing spinor-tensors of $AdS_d$).

We set the AdS radius to $L=1$ and we work in the Poincar\'e patch parameterised as
\be \label{poincare-z}
ds^2 = \frac{1}{z^2} \left( dz^2 + \h_{IJ} dx^I dx^J \right) .
\ee 
We also fix the local frame as
\be
e_\m{}^A = \frac{1}{z}\, \d_\m{}^A \, , \quad \o_I{}^{zJ} = \frac{1}{z}\, \d_I{}^J \, , \quad \o_z{}^{\m\n} = \o_I{}^{JK} = 0 \, ,
\ee
where we take advantage of the form of the vielbein to identify ``flat'' and ``curved'' indices.
In these coordinates the spatial boundary is at $z \to 0$. All results can be easily translated in the coordinates \eqref{AdS} used in the main body of the text, in which the boundary is at $r \to \infty$. We denote by capital Latin indices all directions transverse to the radial one (including time).

\subsection{Falloff of the solutions of the free equations of motion}\label{sec:eom-fermi}

We wish to study the solutions of the Fang-Fronsdal equation on a constant-curvature background of dimension $d$ \cite{fronsdal-AdS} which, in the index-free notation of sect.~\ref{sec:arbitrary}, reads
\be \label{fang-fronsdal}
i \left( \slashed{D} \psi - s\, D \slashed{\psi} + \frac{d+2(s-2)}{2}\, \psi + \frac{s}{2}\, \g\, \slashed{\psi} \right) = 0 \, .
\ee
To this end it is convenient to partially fix the gauge freedom \eqref{gauge-fermi} by setting to zero the $\g$-trace of the field (see \cite{dWF} or sect.~2.2 of the review \cite{modave1} for a discussion of this partial gauge fixing in flat space). This leads to the system of equations
\begin{subequations} \label{fierz-fermi}
\begin{align} 
& i \left( \slashed{D} + \frac{d+2(s-2)}{2} \right) \psi = 0 \, , \label{eom-fermi} \\
& \slashed{\psi} = 0 \, . \label{gamma}
\end{align}
\end{subequations}
These conditions also imply that the divergence of the field vanishes: taking the $\g$-trace of the first equation one indeed obtains
\be \label{div-fermi}
0 = \g^\m \slashed{D} \psi_{\m} = - \slashed{D} \slashed{\psi} + 2\, D\cdot \psi \, ,
\ee
that implies $D\cdot \psi = 0$ thanks to the second equation.
Imposing \eqref{gamma} does not fix completely the gauge freedom: eqs.~\eqref{fierz-fermi} admit a residual gauge symmetry with parameters constrained as
\begin{subequations} \label{fierz-gauge-fermi}
\begin{align}
& \! \left( \slashed{D} + \frac{d+2(s-1)}{2} \right) \e = 0 \, , \label{gauge-eom-fermi} \\
& D\cdot \e = 0 \, , \label{gauge-div-fermi} \\
& \slashed{\e} = 0 \, , \label{gauge-gamma}
\end{align}
\end{subequations}
where the cancellation of the divergence follows from the other two conditions as above.\footnote{Eqs.~\eqref{fierz-gauge-fermi} manifestly guarantee that gauge transformations of the form \eqref{gauge-fermi} preserve the $\gamma$-trace constraint \eqref{gamma}. The divergence constraint \eqref{div-fermi} is also preserved thanks to
\[
\d\, D\cdot \psi = \left( \Box + \frac{1}{2} \slashed{D} - \frac{(s-1)(2d+2s-5)}{2} \right) \e = \left( \slashed{D} - \frac{d+2s-3}{2} \right) \! \left( \slashed{D} + \frac{d+2(s-1)}{2} \right) \e \, .
\]
}

To analyse the falloff at $z \to 0$ of the solutions of \eqref{fierz-fermi} one has to treat separately field components with a different number of indices along the $z$ direction. We denote them as
\be
\psi_{z_n I_{s-n}} \equiv \psi_{z \cdots z I_1 \cdots I_{s-n}} \, .
\ee
The $\g$-trace constraint \eqref{gamma} then gives
\be \label{gamma-exp}
\hat{\g}^z\, \psi_{z\m_{s-1}} + \hat{\g}\cdot \psi_{\m_{s-1}} = 0 \, ,
\ee
where here and below contractions only involve indices transverse to $z$.
Using \eqref{gamma-exp}, the components of the equation of motion \eqref{eom-fermi} read
\be \label{eom-exp-fermi}
\begin{split}
& \hat{\g}^z \left( z\,\pr_z - \frac{d-2(s-n)-1}{2} \right) \psi_{z_n I_{s-n}} + \frac{d+2(s-2)}{2}\, \psi_{z_n I_{s-n}}  \\
& + z\, \hat{\g}^J \pr_J \psi_{z_n I_{s-n}} - (s-n)\, \hat{\g}_I \psi_{z_{n+1}I_{s-n-1}}  = 0 \, ,
\end{split}
\ee
where here and in the rest of this appendix repeated covariant or contravariant indices denote a symmetrisation.
To analyse these equations it is convenient to begin from the divergence constraint they imply, 
\be \label{div-exp-fermi}
\left( z\,\pr_z - d + \frac{3}{2} \right) \psi_{z\m_{s-1}} + z\, \prd \psi_{\m_{s-1}} = 0 \, ,
\ee
which entails $\psi_{z_n I_{s-n}} \sim z^{\D + n}$. Even if the equations are of first order, two values of $\Delta$ are admissible due to the dependence on $\hat{\g}^z$ in \eqref{eom-exp-fermi}. Asymptotically one can indeed split each component of the field as
\be
\psi_{z_n I_{s-n}} = \psi^{+}_{z_n I_{s-n}} + \psi^{-}_{z_n I_{s-n}} \, ,
\ee
where the $\psi^\pm$ are eigenvectors of $\hat{\g}^z$, i.e.
\be \label{eigenvectors}
\hat{\g}^z \psi^{\pm}_{z_n I_{s-n}} = \mp\, \psi^{\pm}_{z_n I_{s-n}} \, .
\ee
Substituting this ansatz in \eqref{eom-exp-fermi}, the terms in the second line are subleading for $z \to 0$ and the first line vanishes provided that
\be \label{fall-off-fermi-field}
\psi^{\pm}_{z\cdots z\, I_1 \cdots I_{s-n}} \sim z^{\D_\pm+\,n} \quad \textrm{with} \ \
\left\{
\begin{array}{l}
\D_+ = d-\frac{5}{2} \\[5pt]
\D_- = \frac{3}{2}-2s
\end{array}
\right. .
\ee
This implies that asymptotically one has to force a projection as already noticed for $s = 3/2$ and $d= 4$ \cite{HT} (see also \cite{bnd_spin_5-2,bnd_spin_5-2_2} for the extension to arbitrary $d$ and \cite{AdS/CFT-spinors,boundary-dirac} for $s = 1/2$). A comparison with the fall-off conditions for Bose fields recalled in (C.9) of \cite{charges-bose} shows that
\be
\D^{Fermi}_{\pm} = \D^{Bose}_{\pm} \pm \frac{1}{2} \, ,
\ee
while for $s = 0$ one recovers the asymptotic behaviour of a Dirac fermion of mass \mbox{$m^2 = - 2(d-3)$}.

\subsection{Residual gauge symmetry}\label{sec:gauge-fermi}

The fall-off conditions of the parameters of the residual gauge symmetry are fixed by eqs.~\eqref{fierz-gauge-fermi}. The divergence and trace constraints give
\begin{align}
\left( z\,\pr_z - d + \frac{3}{2} \right) \e_{z\m_{s-2}} + z\, \prd \e_{\m_{s-2}} & = 0 \, , \\
\hat{\g}^z\, \e_{z\m_{s-2}} + \hat{\g}\cdot \e_{\m_{s-2}} & = 0 \, ,
\end{align}
and the first condition implies $\e_{z_n j_{s-n-1}} \sim z^{\Th + n}$. By using these identities in \eqref{gauge-eom-fermi} one obtains
\be \label{gauge-exp-fermi}
\begin{split}
& \hat{\g}^z \left( z\,\pr_z - \frac{d-2(s-n)+1}{2} \right) \e_{z_n I_{s-n-1}} + \frac{d+2(s-1)}{2}\, \e_{z_n I_{s-n-1}}  \\
& + z\, \hat{\g}^J \pr_J \e_{z_n I_{s-n-1}} - (s-n-1)\, \hat{\g}_I \e_{z_{n+1}I_{s-n-2}}  = 0 \, .
\end{split}
\ee
This equation has the same form as \eqref{eom-exp-fermi}, apart from the shift $s \to s-1$ and a modification in the mass terms. As a result, by decomposing the gauge parameters as $\e = \e^+ + \e^-$ with 
\be \label{eigenvectors2}
\g^z \e^{\pm}{}_{z_n I_{s-n-1}} = \mp\, \e^{\pm}{}_{z_n I_{s-n-1}}
\ee
and following the same steps as above one obtains
\be \label{x-cov-fermi}
\e^\pm_{z\cdots z\, I_1 \cdots I_{s-n-1}} \sim z^{\Theta_\pm+\,n} \quad \textrm{with} \ \
\left\{
\begin{array}{l}
\Theta_+ = d-\frac{1}{2} \\[5pt]
\Theta_- = \frac{3}{2}-2s
\end{array}
\right. .
\ee
A comparison with the fall-off conditions in (C.12) of \cite{charges-bose} shows that
\be
\Th^{Fermi}_{\pm} = \Th^{Bose}_{\pm} \pm \frac{1}{2}
\ee
also for gauge parameters.

One can compare these results with the conditions satisfied by a gauge transformation preserving the AdS background, for which
\be \label{killing-fermi}
\d \psi = s \left( D \e + \frac{1}{2}\, \g\, \e \right) = 0 \, ,
\qquad \slashed{\e} = 0 \, .
\ee
These constraints also imply $D\cdot\e =0$. Expanding \eqref{killing-fermi} one obtains
\be \label{killing-vs-boundary}
\left( z\,\pr_z + (2s-n-2) + \frac{1}{2}\, \hat{\g}^z \right) \e_{z_nI_{s-n-1}} = \cO(z^{\Th + n + 1}) \, .
\ee
This equation is analysed more in detail in sect.~\ref{sec:bnd}; here it is worth noting that the solutions in the $\Theta_-$ branch of \eqref{x-cov-fermi} also solve \eqref{killing-vs-boundary}.

\subsection{Initial data at the boundary}\label{sec:leading-fermi}

In this subsection we display the constraints on the initial data at the boundary imposed by the equations of motion and the $\g$-trace constraint, and how the number of independent components is further reduced by the residual gauge symmetry. First of all, note that the solutions of \eqref{fierz-fermi} are generically of the form
\be
\psi_{z_mI_{s-m}} = \sum_{n=0}^\infty z^{\D_+ +m+n } q^{(m,n)}_{I_{s-m}}(x^k) 
\quad \textrm{or} \quad
\psi_{z_mI_{s-m}} = \sum_{n=0}^\infty z^{\D_- +m+n } \rho^{(m,n)}_{I_{s-m}}(x^k) \, ,
\ee 
where all spinor-tensors in the series have a definite (alternating) chirality:\footnote{The components of e.g.\ the $q^{(m,n)}_{I_{s-n}}$ can be considered as spinors defined on the $(d-1)$-dimensional boundary of AdS. When $d$ is odd, \eqref{proj-q-r} is a chirality projection. When $d$ is even, a priori the boundary values of $\psi_{IJ}$ would be collected in a couple of Dirac spinors and \eqref{proj-q-r} selects one of them.}
\be \label{proj-q-r}
\hat{\g}^z q^{(m,n)} = (-1)^{n+1} q^{(m,n)} \, , \qquad 
\hat{\g}^z \r^{(m,n)} = (-1)^n \r^{(m,n)} \, .
\ee
The $\g$-trace constraint \eqref{gamma-exp} then allows one to solve all components $\psi_{z_nI_{s-n}}$ with $n \geq 1$ in terms of the purely transverse one, $\psi_{I_s}$. The equation of motion \eqref{eom-exp-fermi} finally fixes the subleading components of $\psi_{I_s}$ in terms of the leading one.

Within the admissible fall-off conditions, the $\D_+$ branch is the one which is relevant for the analysis of surface charges. We denote its leading contributions in $\psi_{I_s}$ as 
\be \label{vev}
\psi_{I_s} = z^{\D_+} \cQ^-_{I_s}(x^K) + z^{\D_+ + 1} \cQ^+_{I_s}(x^K) + \cO(z^{\D_+ + 2}) \, , \qquad 
( 1 \mp \hat{\g}^z ) \cQ^\pm_{I_s} = 0 \, .
\ee
For completeness, we shall also analyse the constraints imposed on the leading contributions in the $\D_-$ branch, denoted as
\be \label{source}
\psi_{I_s} = z^{\D_-} \Psi^+_{I_s}(x^K) + z^{\D_- + 1} \Psi^-_{I_s}(x^J) + \cO(z^{\D_- + 2}) \, , \qquad 
( 1 \mp \hat{\g}^z ) \Psi^\pm_{I_s} = 0 \, .
\ee
The spinor-tensors $\cQ^-_{I_s}$ and $\Psi^+_{I_s}$ are boundary fields of opposite chirality (or, when the dimension of the boundary is odd, Dirac fields with different eigenvalues of $\hat{\g}^z$) of conformal dimensions, respectively, $\D_c = d+s-\frac{5}{2}$ and $\Delta_s = \frac{3}{2}-s$. They thus correspond to the fermionic \emph{conserved currents} and \emph{shadows fields} of \cite{shadows-fermi}.\footnote{If one performs a dilatation $x^\m \to \l x^\m$ then $\psi_{I'_s} = \l^{-s} \psi_{I_s}$, while on the right-hand side of \eqref{vev} or \eqref{source} one has $z'^{\D_\pm} = \l^{\D_\pm} z^{\D_\pm}$. As a result, both $\cQ$ and $\Psi$ must transform as $\cQ_{I'_s} = \l^{- ( \Delta_+ + s )} \cQ_{I_s}$, from where one reads the conformal dimensions. To compare them with eqs.~(5.8) and (6.6) of \cite{shadows-fermi}, consider that $d_{here} = (d+1)_{there}$.} 

Combining the e.o.m.\ (or, equivalently, the divergence constraint) and the $\g$-trace constraint gives
\be
\pr^J \cQ^-_{JI_{s-1}} = 0 \, , \qquad
\hat{\g}^J \cQ^-_{JI_{s-1}} = \h^{JK}\cQ^+_{JKI_{s-2}} = 0 \, .
\ee
Eq.~\eqref{eom-exp-fermi} also allows one to fix the $\g$-traceless component of $\cQ^+$ as
\be
\left(\cQ^+_{I_s}\right)^{Tr} \equiv \cQ^+_{I_s} - \frac{s}{d+2s-3}\, \hat{\g}^{\phantom{+}}_{I}\! \slashed{\cQ}^+_{I_{s-1}} = - \frac{1}{d+2s-3}\, \slashed{\pr} \cQ^-_{I_s} \, . 
\ee
The $\g$-trace of $\cQ^+$ remains free as the divergenceless part of $\cQ^-$. In the $\D_-$ branch the $\g$-trace constraint similarly imposes
\be
 \hat{\g}^J \Psi^+_{JI_{s-1}} = \h^{JK}\Psi^-_{JKI_{s-2}} =0 \, .
\ee
The full $\g$-traceless $\Psi^+$ remains instead unconstrained, while all $\Psi^-$ is now fixed as
\be
\Psi^-_{I_s} = - \frac{1}{d+2s-5}\left( \slashed{\pr} \Psi^+_{I_s} - \frac{s}{d+2s-4}\, \hat{\g}^{\phantom{+}}_{I}\! \prd \Psi^+_{I_{s-1}} \right) .
\ee
Note that, in analogy with Bose field \cite{charges-bose}, the total number of independent data that asymptotically can be chosen arbitrarily\footnote{Of course, regularity  in the bulk should  fix the vev in terms of the source.  We do not discuss this issue here as we focus on the asymptotic behaviour of the theory. Integration in the bulk necessitates the full theory beyond the linear terms.} is the same in both branches, even if they are distributed in different ways in \eqref{vev} and \eqref{source}.

The number of independent initial data is further reduced by the residual gauge symmetry. The components of the field vary as
\be
\begin{split}
z\,\d \psi_{z_n I_{s-n}} & = n \left( z\,\pr_z + 2s - n - 1 + \frac{\hat{\g}^z}{2} \right) \e_{z_{n-1}I_{s-n}} + \frac{s-n}{2}\, \hat{\g}_I\! \left( 1 - \hat{\g}^z \right) \e_{z_{n}I_{s-n-1}} \\
& + (s-n) \left( z\,\pr_I \e_{z_nI_{s-n-1}} - (s-n-1) \h_{II} \e_{z_{n+1}I_{s-n-2}} \right) .
\end{split}
\ee
Gauge transformations generated by
\be
\e_{z_n I_{s-n-1}} = z^{\Th_+ + n}\, \xi^{(n)}_{I_{s-n-1}}(x^K) + \cO(z^{\Th_+ + n + 1}) \, , \qquad
(1+\hat{\g}^z)\, \xi^{(n)}_{I_{s-n-1}} = 0 \, ,  
\ee
naturally act on the $\D_+$ branch of solutions of the e.o.m.: they allow to set to zero the $\psi_{z_nI_{s-n}}$ components with $n \geq 1$ (and therefore also $\slashed{\cQ}^+$), while they leave $\cQ^-$ invariant. On the other hand, gauge transformations generated by
\be
\e_{z_n I_{s-n-1}} = z^{\Th_- + n}\, \varepsilon^{(n)}_{I_{s-n-1}}(x^K) + \cO(z^{\Th_- + n + 1}) \, , \qquad
(1-\hat{\g}^z)\, \varepsilon^{(n)}_{I_{s-n-1}} = 0 \, , 
\ee
naturally act on the $\D_-$ branch and they affect the leading contribution as
\be
\d \Psi^+_{I_s} = s\, \pr_I \ve_{I_{s-1}} - \frac{s}{d+2s-3} \left( \hat{\g}_I \slashed{\pr} \ve_{I_{s-1}} + (s-1) \h_{II} \prd \ve_{I_{s-2}} \right) ,
\ee
where we defined $\ve \equiv \ve^{(0)}$ and the variation is $\g$-traceless as it should.
This gauge freedom reduces the number of independent components $\Psi^+$ such that it becomes identical to that of the conserved current $\cQ^-$. It also leaves the coupling $i\! \left( \bar{\Psi}^+ \cQ^- + \bar{\cQ}^- \Psi^+ \right)$ invariant.

\section{Conformal Killing spinor-tensors}\label{app:conformal-killing}

In this appendix we present the general solution of the conformal Killing equations \eqref{conformal-killing} for spinor-vectors in $d > 3$ (for $d=3$ see sect.~\ref{sec:3d}). We then estimate the number of independent solutions of the conformal Killing equations \eqref{generic-conf-kill} for spinor-tensors of arbitrary rank. We conclude by proving the identities that we used in sect.~\ref{sec:example-symm} to show that asymptotic symmetries are generated by parameters satisfying only the conditions \eqref{final-gauge} and \eqref{conformal-killing}. We also display recursion relations that, for arbitrary values of the rank, allow one to express all subleading components of the parameters generating asymptotic symmetries in terms of the leading ones.

\subsection{Conformal Killing spinor-vectors}\label{app:spinor-vectors}

Ignoring the spinorial index, eq.~\eqref{conf2-} has the same form as the conformal Killing vector equation in Minkowski space. It is therefore natural to consider the ansatz
\be \label{gen-sol-kill}
\z^+_I = v^+_I - \left( x_J \hat{\g}^J \right) v^-_I \, , \qquad
\z^-_I = v^-_I \, ,
\ee
where the $v^\pm_I$ have the same dependence on $x^I$ as conformal Killing vectors:
\be \label{v_i}
v^\pm_I \equiv a^\pm_I + \o^\pm_{IJ}\, x^J + b^\pm\,x_I + c^\pm_J \left( 2 x_I x^J - x^2 \d_I{}^J \right) , 
\qquad
(1 \mp \hat{\g}^r) v^\pm_I = 0 \, .
\ee
Eqs.~\eqref{gen-sol-kill} generalise the general solution of the conformal Killing spinor equations \cite{AdS-killing-spinors} and, indeed, they solve \eqref{conf2+} and \eqref{conf2-} for constant spinor-tensors $a^\pm_I$, $\o^\pm_{IJ}$, $b^\pm$ and $c^\pm_I$ only subjected to the chirality projections inherited from $\z^\pm_I$. The $\g$-trace conditions \eqref{constr2+} and \eqref{constr2-} impose relations between these spinor-tensors, that one can conveniently analyse by decomposing them in $\g$-traceless components. For instance, the $\g$-traceless projections of $a^\pm_I$ and of the antisymmetric $\o^\pm_{IJ}$ are defined as
\begin{subequations}
\begin{align}
\hat{a}^\pm_I & \equiv a^\pm_I - \frac{1}{d-1}\, \hat{\g}^{\phantom{\pm}}_I \!\slashed{a}^\pm \, , \\
\hat{\o}^\pm_{IJ} & \equiv \o^\pm_{IJ} + \frac{2}{d-3}\, \hat{\g}^{\phantom{\pm}}_{[I} \hat{\g}^K \o^\pm_{J]K} - \frac{1}{(d-2)(d-3)}\, \hat{\g}_{IJ} \hat{\g}^{KL} \o_{KL} \, .
\end{align}
\end{subequations}
The constraints relate the $\g$-traces of $a^\pm_I$, $\o^\pm_{IJ}$ and $c^\pm_I$ to other spinor-tensors in \eqref{v_i}, such that the general solution of the full system of equations \eqref{conformal-killing} is given by
\be
\begin{split}
\z^+_I & = \hat{a}^+_I + x^J\! \left\{ \hat{\o}^+_{IJ} - \hat{\g}^{\phantom{+}}_{(I} \hat{a}^-_{J)} + \frac{d+1}{d-3}\, \hat{\g}^{\phantom{+}}_{[I} \hat{a}^-_{J]} + \frac{d}{d-1}\, \h_{IJ} b^+ - \frac{d}{(d-1)(d-2)}\, \hat{\g}_{IJ} b^+ \right\} \\
& + x^Jx^K \left\{ \frac{2(d-2)}{d-3}\, \h^{\phantom{+}}_{I(J} \hat{c}^+_{K)} - \frac{d-1}{d-3}\, \h^{\phantom{+}}_{JK} \hat{c}^+_{I} - \frac{2}{d-3}\, \hat{\g}^{\phantom{+}}_{I(J} \hat{c}^+_{K)} + \hat{\g}^{\phantom{-}}_{(J} \hat{\o}^-_{K)I} \right. \\
& \left. - \frac{d}{d-2} \left( \h_{I(J} \hat{\g}_{K)} b^- - \frac{1}{d-1}\, \h_{JK} \hat{\g}_I b^- \right) \right\} + x^Jx^Kx^L \left\{ \h^{\phantom{-}}_{(JK} \hat{\g}^{\phantom{-}}_{L)} \hat{c}^-_I - 2\, \h^{\phantom{-}}_{I(J} \hat{\g}^{\phantom{-}}_K \hat{c}^-_{L)} \right\}
\end{split}
\ee
and
\be
\begin{split}
\z^-_I & = \hat{a}^-_I - \frac{1}{d-1}\, \hat{\g}_I b^+ + x^J \left\{ \hat{\o}^-_{IJ} - \frac{4}{d-3}\, \hat{\g}^{\phantom{+}}_{[I} \hat{c}^+_{J]} + \h_{IJ} b^- + \frac{2}{(d-1)(d-2)}\, \hat{\g}_{IJ} b^- \right\} \\
& + x^Jx^K \left\{ 2\,\h^{\phantom{-}}_{I(J} \hat{c}^-_{K)} - \h^{\phantom{-}}_{JK} \hat{c}^-_{I} \right\} .
\end{split}
\ee

\subsection{Comments on arbitrary rank}\label{app:conformal-general}

In the previous subsection we have seen that, in the rank-1 case, the general solution of the conformal Killing equations \eqref{generic-conf-kill} in $d-1$ dimensions depends on the integrations constants collected in the $\g$-traceless spinor-tensors $\hat{a}^\pm_I$, $\hat{\o}^\pm_{IJ}$, $b^\pm$ and $\hat{c}^\pm_I$. Hence, in analogy with what happens for ``bosonic'' conformal Killing tensors \cite{algebra}, there are as many integration constants as independent $\g$-traceless Killing spinor-tensors of a Minkowski space of dimensions $d$. The equations
\be
\d_{(flat)} \psi_{\m\n} = 2\,\pr_{(\m} \e_{\n)} = 0 \, , \qquad
\g^\m \e_\m = 0
\ee
are indeed solved by
\be \label{sol-flat-killing}
\e_\m = A_\m + B_{\m\n} x^\n \, , \qquad 
\g^\m A_\m = \g^\m B_{\m\n} = B_{(\m\n)} = 0 \, ,
\ee
and the number of independent components of the constants $A_\m$ and $B_{\m\n}$ in $d$ dimensions equates that of $\hat{a}^\pm_I$, $\hat{\o}^\pm_{IJ}$, $b^\pm$ and $\hat{c}^\pm_I$ in $d-1$ dimensions. This indicates that the number of independent $\g$-traceless Killing spinor-tensors on AdS and Minkowski backgrounds is the same (at least up to rank 1).
 
The pattern of independent spinor-tensors entering \eqref{v_i} can be understood from the branching rules for representations of the orthogonal group (see e.g.\ \S~8.8.A of \cite{Barut}). Denoting a Young tableau with $s$ boxes in the first row and $k$ boxes in the second by $\{s,k\}$, a $\g$-traceless $\{s,k\}$-projected spinor-tensor in $d+1$ dimensions decomposes in a sum of two-row projected spinor-tensors in $d$ dimensions as\footnote{This rule can be also checked by considering that the number of components of a $\{s,k\}$-projected $\g$-traceless Dirac spinor-tensor in $d$ dimensions is (see e.g.~(A.39) of \cite{review-mixed})
\[
\textrm{dim}_{O(d)}\{s,k\} = 2^{\left[\frac{d}{2}\right]}\frac{(s-k+1)}{(s+1)!k!}\, \frac{(d+s-3)!(d+k-4)!}{(d-2)!(d-4)!}\, (d+s+k-2) \, .
\]}
\be \label{branching}
\{s,k\}_{d} = \sum_{r=k}^s \sum_{l=0}^k\, n(d) \{r,l\}_{d-1} \, ,
\ee
where the multiplicity factor $n(d)$ is equal to 1 when $d$ is odd and to $2$ when $d$ is even. Applying this rule to $A_\m$ and $B_{\m\n}$, one recovers the spinor-tensors entering \eqref{v_i}. When $d$ is odd, the $\pm$ doubling in \eqref{v_i} allows to reproduce the components of a Dirac $\e_\m$ from two sets of Weyl spinor-tensors. When $d$ is even, the doubling accounts for the factor $n(d)$ in \eqref{branching}.

A full derivation of the solutions of the conformal Killing equations \eqref{generic-conf-kill} will be given elsewhere (see also \cite{superconformal} for related work based on superspace techniques). Here we assume that the pattern emerged in the rank $0$ and $1$ examples extends to arbitrary values of the rank. Accordingly, we assume that, for $d > 3$, the number of independent $\g$-traceless Killing spinor-tensors on AdS and Minkowski backgrounds is the same. In the limit $L \to \infty$ the solutions of the Killing equations \eqref{killing-ads-gen} are given by
\be \label{killing-flat}
\e_{\m_1 \cdots \m_s} = \sum_{k=0}^s \, A_{\m_1 \cdots \m_s | \n_1 \cdots \n_k} x^{\n_1} \cdots x^{\n_k} \, , 
\quad
\g^\r A_{\r\,\m_2 \cdots \m_{s} | \n_1 \cdots \n_k} = A_{(\m_1 \cdots \m_s | \m_{s+1}) \n_1 \cdots \n_{k-1}} = 0 \, .
\ee
Their number is therefore equal to the number of components of a $\g$-traceless (Weyl) spinor-tensor in $d+1$ dimensions with the symmetries of a rectangular $\{s,s\}$ Young tableau, that is to
\be
\textrm{dim}_{O(d+1)}\{s,s\} = 2^{\left[\frac{d}{2}\right]} \frac{(d+s-2)!(d+s-3)!(d+2s-1)}{s!(s+1)!(d-1)!(d-3)!}\,  \, .
\ee

\subsection{Independent conditions on asymptotic symmetries}\label{app:independent}

\subsubsection*{Identities involving conformal Killing spinor-vectors}

In order to verify that the conditions \eqref{final-gauge} and \eqref{conformal-killing} on the gauge parameter $\e^\m$ fully characterise asymptotic symmetries for spin-$5/2$ fields, one has to prove that \eqref{diff-constr} holds and that the second line in \eqref{dpsi-ij-exp} vanishes. This requires
\begin{align}
\frac{d-1}{d}\, \slashed{\pr} \prd \z^+ + (d+1)\, \prd\z^- & = 0 \, , \label{extra1} \\
\slashed{\pr} \prd\z^- & = 0 \, , \label{extra2} \\
\frac{d-1}{d}\, \pr_I \pr_J \prd\z^+ + 2\, \hat{\g}_{(I} \pr_{J)} \prd \z^- & = 0 \, , \label{extra3} \\
\pr_I \pr_J \prd \z^- & = 0 \, . \label{extra4}
\end{align}

We wish to prove that the identities above follow from \eqref{conformal-killing}. One can obtain scalars from these equations only by computing a double divergence or a divergence and a $\g$-trace (since the Killing equations are traceless). Eliminating $\slashed{\z}^\pm$ via \eqref{constr2+} and \eqref{constr2-}, the double divergences of \eqref{conf2+} and \eqref{conf2-} become, respectively,
\begin{align}
\frac{d-1}{d}\, \Box \prd\z^+ + \slashed{\pr} \prd \z^- & = 0 \, , \label{dd+} \\
\frac{d-2}{d-1} \, \Box \prd \z^- & = 0 \, . \label{dd-}
\end{align}
Computing a divergence and a $\g$-trace one obtains instead
\begin{align}
\frac{d-1}{d}\, \slashed{\pr} \prd\z^+ + (d+1) \prd \z^- & = 0 \, , \label{dg+} \\
\frac{d-1}{d}\, \Box \prd \z^+ - (d-3) \slashed{\pr} \prd \z^- & = 0 \, . \label{dg-}
\end{align}
Eq.~\eqref{dg+} directly shows that \eqref{extra1} is not independent from \eqref{conformal-killing}. Moreover, combining \eqref{dd+} and \eqref{dg-}, for $d > 2$ one obtains
\be \label{boxes}
\Box \prd\z^+ = \Box \prd \z^- = \slashed{\pr} \prd \z^- = 0 \, ,
\ee
so that \eqref{extra2} is not independent as well. All in all, this implies that the $\g$-trace constraint \eqref{diff-constr} is satisfied when the conformal Killing equations \eqref{conformal-killing} hold.

One can prove \eqref{conf2-} $\Rightarrow$ \eqref{extra4} by acting with a gradient on \eqref{conf2-} and manipulating the result as in Appendix~D of \cite{charges-bose}:
\be
\begin{split}
0 & = 2\,\pr_K\! \left( \pr_{(I} \z_{J)}{}^{\!\!-} - \frac{1}{d-1}\,\h_{IJ}\, \prd \z^- \right) \\
& = 3\! \left( \pr_{(I} \pr_J \z_{K)}{}^{\!\!-} - \frac{2}{d-1}\,\h_{(IJ}\pr_{K)} \prd \z^- \right) - \pr_I \pr_J \z_{K}{}^{\!\!-} + \frac{4}{d-1}\, \h_{K(I} \pr_{J)} \prd \z^- \label{trick1} \\
& = 3\! \left( \pr_{(I} \pr_J \z_{K)}{}^{\!\!-} - \frac{1}{d-1}\,\h_{(IJ}\pr_{K)} \prd \z^- \right) - \pr_I \pr_J \z_{K}{}^{\!\!-} + \frac{2}{d-1}\, \h_{K(I} \pr_{J)} \prd \z^- \\
& \phantom{=}\, - \frac{1}{d-1}\, \h_{IJ} \pr_K \prd \z^- \, .
\end{split}
\ee
The terms between parentheses in the last step vanish independently because they are the symmetrisation of the first line. Contracting the other three terms with $\pr^K$ one then obtains (for $d > 1$)
\be
(d-3) \pr_I \pr_J \prd \z^- + \h_{IJ} \Box \prd \z^- = 0 \, .
\ee
The last addendum vanishes thanks to \eqref{dd-} (double divergence of \eqref{conf2-}), so that \eqref{extra4} is satisfied when $d > 3$. In $d =3$ the missing cancellation originates the variation of surface charges discussed at the end of sect.~\ref{sec:example-symm}.

One can prove that \eqref{extra2} is not independent in a similar way. First of all, let us manipulate \eqref{conf2+} (here combined with \eqref{constr2-}) as in \eqref{trick1}:
\be \label{trick2}
\begin{split}
0 & = 2\,\pr_K \!\left( \pr_{(I} \z_{J)}{}^{\!\!+} - \frac{1}{d}\,\h_{IJ}\, \prd \z^+ + \hat{\g}_{(I} \z_{J)}{}^{\!\!-}  \right) \\
& = 3\! \left( \pr_{(I} \pr_J \z_{K)}{}^{\!\!+} - \frac{2}{d}\,\h_{(IJ}\pr_{K)} \prd \z^+ + 2\, \hat{\g}_{(I} \pr_J \z_{K)}{}^{\!\!-} \right) - \pr_I \pr_J \z_{K}{}^{\!\!+} + \frac{4}{d}\, \h_{K(I} \pr_{J)} \prd \z^+ \\
& \phantom{=}\, - 2\, \hat{\g}_{(I} \pr_{J)} \z_{K}{}^{\!\!-} - 2\, \hat{\g}_{K} \pr_{(I} \z_{J)}{}^{\!\!-} \\
& = 3\! \left( \pr_{(I} \pr_J \z_{K)}{}^{\!\!+} - \frac{1}{d}\,\h_{(IJ}\pr_{K)} \prd \z^+ + \hat{\g}_{(I} \pr_J \z_{K)}{}^{\!\!-} \right) - \pr_I \pr_J \z_{K}{}^{\!\!+} + \frac{2}{d}\, \h_{K(I} \pr_{J)} \prd \z^+ \\
& \phantom{=}\, - \frac{1}{d}\, \h_{IJ} \pr_{K} \prd \z^+ 
+ \hat{\g}_{(I|} \pr_{K} \z_{|J)}{}^{\!\!-} - \hat{\g}_{(I} \pr_{J)} \z_{K}{}^{\!\!-} - \hat{\g}_{K} \pr_{(I} \z_{J)}{}^{\!\!-} \, .
\end{split}
\ee
The terms between parentheses in the last step vanish because they are the symmetrisation of the first line. The remaining contributions thus yield another vanishing combination, whose contraction with $\pr^K$ gives
\be \label{trick3}
\frac{d-2}{d}\, \pr_I \pr_J \prd \z^+ + \hat{\g}_{(I} 
\pr_{J)} \prd \z^- - \Box \hat{\g}_{(I} \z_{J)}{}^{\!\!-} + \slashed{\pr} \pr_{(I} \z_{J)}{}^{\!\!-} = 0 \, .
\ee
One can show that $\slashed{\pr} \pr_{(I} \z_{J)}{}^{\!\!-}$ vanishes using first \eqref{conf2-} and then \eqref{boxes}. The term $\Box \hat{\g}_{(I} \z_{J)}{}^{\!\!-}$ is instead proportional to the second contribution, since the divergence of \eqref{conf2-} implies 
\be
\Box \z_{I}{}^{\!\!-} = - \frac{d-3}{d-1}\, \pr_I \prd\z^- \, .
\ee
All in all, \eqref{trick3} becomes
\be
\frac{d-2}{d-1} \left( \frac{d-1}{d}\, \pr_I\pr_J \prd\z^+ + 2\, \hat{\g}_{(I} \pr_{J)} \prd \z^- \right) = 0 \, ,
\ee
thus completing the proof that \eqref{extra3} is not independent from \eqref{conformal-killing}. Note that --~in contrast with the proof of \eqref{extra4}~-- this is true also in $d = 3$, as it is necessary to obtain a variation of $\psi_{IJ}$ satisfying the boundary conditions of sect.~\ref{sec:example-bnd}.

\subsubsection*{Asymptotic Killing spinor-tensors}

In order to show that asymptotic symmetries are generated by gauge parameters of the form \eqref{gauge0-gen} that are fully characterised by $\z^\pm$, one should express the spinor-tensors $\a_k$ and $\b_k$ in terms of the former. This can be done by imposing the cancellation of the variations \eqref{varpsi-gen}. The first variation vanishes provided that 
\begin{subequations} \label{firstset}
\begin{align}
& \pr \a_k + \g\, \b_k + (s-1)\, \h \, \slashed{\b}{}_k = 0 \, , \label{eq1} \\[5pt]
& \pr \b_{k-1} - (s-1)\, \h \, \slashed{\a}{}_k = 0 \, , \label{eq2}
\end{align}
\end{subequations}
while $\d \psi_1$ vanishes provided that
\begin{subequations}
\begin{align}
& 2k\,\b_k + (s-1) \left( \pr \slashed{\a}{}_k + (s-2)\, \h\, \b^{\,\pe}_k \right) = 0 \, , \label{eq3} \\[5pt]
& 2k\,\a_k - (s-1) \left( \pr \slashed{\b}{}_{k-1} - \g\, \slashed{\a}{}_k - (s-2)\, \h\, \a'_k \right) = 0 \label{eq4} \, .
\end{align}
\end{subequations}
The last two equations allow one to fix recursively all $\a_k$ and $\b_k$ in terms of $\a_0 = \z^+$ and $\b_0 = \z^-$ (considering also that the $\g$-trace constraint \eqref{gamma-trace-symm} implies $\slashed{\a}{}_k^{[k]} = \b_k^{[k+1]} = 0$). Consistency with \eqref{firstset} is not manifest; yet, in analogy with what we proved above for $s=2$, it must follow from \eqref{generic-conf-kill}. Assuming compatibility of the full system of equations, one can first solve \eqref{eq2} by taking successive traces so as to obtain
\begin{align}
\slashed{\a}_k^{[n]} & = \!\sum_{j=0}^{k-n-1}\! C(n,j)\, \h^j \! \left\{ 2(n+j+1)\, \prd \b_{k-1}^{[n+j]} +  (s-2(n+j+1))\, \pr \b_{k-1}^{[n+j+1]} \right\} , \\
C(n,j) & = \frac{(-1)^j\, n! (s-2n-2)! (d+2(s-n)-2j-7)!!}{2^{j+1} (n+j+1)! (s-2n-2j-2)! (d+2(s-n)-5)!!} \, .
\end{align}
Combining this result with \eqref{eq4} one gets
\be \label{alpha}
\begin{split}
\a_k & = \frac{s-1}{2k} \left\{ \pr \slashed{\b}{}_{k-1} - \frac{1}{2(d+2s-5)}\, \g \left( 2\,\prd \b_{k-1} + (s-2)\, \pr \b^{\,\pe}_{k-1} \right) \right\} \\
& + \sum_{j=1}^{k-1} A(j,k) \left\{ \h^j \left[ 2j\, \prd\slashed{\b}{}_{k-1}^{[j-1]} + \slashed{\pr} \b_{k-1}^{[j]} + (s-2j-1)\, \pr \slashed{\b}{}_{k-1}^{[j]} \right] \right. \\
& \left. - \frac{(2j+1)(s-2j-1)}{2(j+1)(d+2s-2j-5)}\, \h^j \g \left[ 2(j+1)\, \prd \b_{k-1}^{[j]} + (s-2j-2)\, \pr \b^{[j+1]}_{k-1} \right] \right\}
\end{split}
\ee
with
\be
A(j,k) = \frac{(-1)^j (s-1)! (d+2s-2j-5)!!}{k\,2^{j+1} j! (s-2j-1)!(d+2s-5)!!} \, .
\ee
In a similar fashion, \eqref{eq3} gives directly
\begin{align} \label{sigma}
\b_k & = \sum_{j=0}^{k} B(j,k)\,  \h^j \left\{ 2j\, \prd \slashed{\a}{}_k^{[j-1]} + (s-2j-1)\, \pr \slashed{\a}{}_k^{[j]} \right\} \, , \\
B(j,k) & = \frac{(-1)^{j+1}(s-1)!}{2^{j+1}(s-2j-1)! \prod_{l=0}^j \left[ l(d+2s-2l-5) + k \right]} \, .
\end{align}

\end{appendix}



\end{document}